\documentclass[aps,prb,twocolumn,superscriptaddress,footinbib]{revtex4-2}

\usepackage[hidelinks]{hyperref}
\hypersetup{
	colorlinks,
	linkcolor={red},
	citecolor={blue},
	urlcolor={blue}
}
\usepackage{physics}
\usepackage{balance}
\usepackage{amssymb}
\usepackage{suffix}
\usepackage{mathtools}
\usepackage[utf8]{inputenc}
\usepackage{booktabs}
\usepackage{cases}
\usepackage[multiple]{footmisc}
\usepackage{dcolumn}
\usepackage{color,soul}
\usepackage{rotating}
\usepackage{perpage}
\usepackage{siunitx}
\usepackage{xcolor}
\usepackage{soul}
\usepackage{amsmath}
\usepackage{tikz}
\usepackage[T1]{fontenc}
\usepackage{etoolbox}
\usepackage{graphics}
\usepackage{siunitx}
\usepackage{float}	
\usepackage{collref}
\usepackage{multirow}
\usepackage{mathtools}
\usepackage{bm}
\usepackage{url}

\usepackage{tikz}
\usepackage{tikz-3dplot}
\usepackage{accents}

\makeatletter
\newcommand{\doublewidetilde}[1]{{%
		\mathpalette\double@widetilde{#1}%
}}
\newcommand{\double@widetilde}[2]{%
	\sbox\z@{$\m@th#1\widetilde{#2}$}%
	\ht\z@=.9\ht\z@
	\widetilde{\box\z@}%
}
\makeatother

\newcommand{\redd}[1]{\textcolor{black}{#1}}

\usepackage{etoolbox,lipsum}

\begin{document} 
	\title{Slow-phonon control of spin Edelstein effect in Rashba $d$-wave altermagnets}
	
	\author{Mohsen Yarmohammadi}
	\email{mohsen.yarmohammadi@georgetown.edu}
	\address{Department of Physics, Georgetown University, Washington DC 20057, USA}
	\author{Jacob Linder}
	\address{Center for Quantum Spintronics, Department of Physics, Norwegian University of Science and Technology, NO-7491 Trondheim, Norway}
	\author{James K. Freericks}
	\address{Department of Physics, Georgetown University, Washington DC 20057, USA}
	\date{\today}
	\begin{abstract}
		Altermagnets have zero net magnetization yet feature spin-split bands. Here, we investigate how slow lattice vibrations (phonons) influence both the intrinsic and externally induced spin polarizations in two-dimensional $d$-wave altermagnets. For the induced spin polarization, we employ a Rashba continuum model with electron-phonon coupling~(EPC) treated at the static-Holstein level and analyze the spin Edelstein effect using the Kubo linear-response formalism to probe EPC-induced contributions. We find that, \redd{under a specific symmetry-lowering pattern like a piezomagnetically active strain that explicitly breaks the inherent $C_4 \mathcal{T}$ symmetry}, moderate-to-strong EPC progressively suppresses the induced polarization via both intraband and interband channels, with a threshold coupling marking the onset of complete spin Edelstein depolarization. The depolarization arises from a phonon-induced energy renormalization that leads to a complete collapse of the Fermi surface. While (de)polarization can occur even in the Rashba non-altermagnetic phase, it remains isotropic. The presence of altermagnetism makes it anisotropic and breaks the conventional antisymmetry between spin susceptibilities that occurs with pure spin-orbit coupling, rendering the effect highly relevant for spintronic applications. We further investigate how the phonon coupling to the altermagnetic order, Rashba spin-orbit strength, and carrier doping collectively tune the depolarization. Our findings demonstrate that static phononic effects offer a powerful means for on-demand control of spin polarization, enabling reversible switching between spin-polarized and depolarized states—a key functionality for advancing spin logic architectures and optimizing next-generation spintronic devices.
	\end{abstract}
	
	\maketitle
	{\allowdisplaybreaks
		
		\section{Introduction}
		
		Altermagnets are a recently identified class of magnetic materials that combine compensated magnetic order with spin-split electronic bands, despite having no net magnetization~\cite{PhysRevX.12.040501,PhysRevX.12.040002,PhysRevX.12.031042,doi:10.7566/JPSJ.88.123702,PhysRevX.12.011028,PhysRevB.102.014422,PhysRevB.99.184432,PhysRevMaterials.5.014409}. Their unusual properties arise from symmetry-protected spin polarization and a mechanism of time-reversal symmetry breaking that preserves global spatial inversion. This, in turn, leads to momentum-dependent spin splittings and nontrivial spin textures even in the absence of external magnetic fields~\cite{doi:10.1073/pnas.2108924118,PhysRevLett.130.216701,PhysRevLett.132.086701,PhysRevLett.132.176702}. The combination of vanishing macroscopic magnetization and strongly spin-polarized states makes altermagnets attractive for spintronic applications, where minimizing stray fields and energy dissipation is crucial~\cite{OrbitalSpinLocking,PhysRevLett.128.197202,PhysRevLett.129.137201,PhysRevLett.134.106801,PhysRevLett.134.106802,PhysRevB.110.235101,PhysRevLett.131.076003,PhysRevB.108.054511,PhysRevB.108.L060508,PhysRevB.108.184505,PhysRevB.108.224421,Zhang2024,PhysRevB.109.224502,PhysRevB.108.205410}. Among them, $d$-wave altermagnets have been mainly studied and noted to host unconventional current-driven spin phenomena~\cite{Bose2022,PhysRevLett.128.197202,PhysRevLett.129.137201,PhysRevLett.126.127701,doi:10.1126/sciadv.adn0479,PhysRevB.109.144421}.
		
		The capability to control intrinsic spin polarization and to externally induce spin polarization in altermagnets offers a powerful pathway for device integration in spintronics. One promising route is through the Edelstein effect~\cite{edelstein1990spin,PhysRevLett.118.116801,Chakraborty2025,k9p4-tfhd,Salemi2019,Johansson_2024,PhysRevMaterials.5.074407,PhysRevMaterials.6.095001,PhysRevMaterials.6.095001,Salemi2019,golub2025spinorientationelectriccurrent,Hu2025}, where electric fields drive spin polarizations in systems with broken inversion symmetry. Such inversion asymmetry, in turn, can be engineered externally, for example by applying gate potentials that induce Rashba spin-orbit coupling~(RSOC)~\cite{yarmohammadi2025anisotropiclighttailoredrkkyinteraction,PhysRevB.110.054427}. In this way, altermagnets can be electrically tuned to host spin-polarized states on demand, opening pathways toward low-power spin-charge interconversion, gate-controllable spin currents, and functional components for next-generation spintronic devices. Monolayer altermagnets have been shown to exhibit spin polarization when subjected to a perpendicular electric field~\cite{mazin2023inducedmonolayeraltermagnetismmnpsse3}. In parallel, theoretical proposals have uncovered a rich variety of current-driven spin responses: relativistic mechanisms at altermagnetic interfaces~\cite{trama2024nonlinearanomalousedelsteinresponse}, nonrelativistic ones across bulk altermagnets~\cite{golub2025spinorientationelectriccurrent}, $p$-wave anti-altermagnets~\cite{hellenes2024pwavemagnets,Jungwirth2025,Chakraborty2025}, and magnets with chiral spin textures~\cite{Hu2025}.
		
		Although altermagnets offer exciting opportunities for spintronic applications, the impact of lattice vibrations (phonons) on their intrinsic and extrinsic spin polarizations has not yet been fully understood. Electron-phonon coupling (EPC) can renormalize quasiparticles and enable phonon-assisted transport in electronic and magnetic systems~\cite{giustino2016electron,PhysRevB.101.121102,10.1063/5.0140724,D3MH00570D,PhysRevB.110.L060409,PhysRevLett.129.076402,PhysRevLett.99.236802,PhysRevLett.110.046402,PhysRevB.92.085137,PhysRevX.15.021039,Patrick_2014}. Very recently, Iorsh~\cite{6wxh-p4mc} showed that dispersive phonons can mediate reentrant superconductivity in altermagnets. Leraand \textit{et al.}~\cite{g4dl-1ff2} demonstrated that, in the weak EPC regime, the leading superconducting instability is odd in momentum and even in spin, yielding spin-polarized Cooper pairs. Steward \textit{et al.}~\cite{nlpj-1dt5} reported hybrid paramagnon-polaron modes arising from coupling between altermagnetic order and phonons. Moreover, Hodt \textit{et al.}~\cite{hodt2025phononenhancedopticalspinconductivityspinsplitter} found a strong phonon-induced enhancement of finite-frequency spin conductivity, while He \textit{et al.}~\cite{j8bg-xbtp} highlighted the role of electron-phonon scattering in controlling charge-to-spin conversion. 
		
		In typical two-dimensional (2D) Rashba systems, the electronic bandwidth far exceeds the characteristic phonon energy scale~\cite{PhysRevB.81.075306,Xu2023}, implying that phonons evolve much more slowly than electronic degrees of freedom. This separation of timescales underlies many theoretical treatments of spin phenomena in altermagnets, yet remains largely unexplored. Despite the ubiquity of phonons, the influence of slow phonons (low frequency/long period)—common in real materials—on low-energy spin excitations and the spin Edelstein effect in Rashba altermagnets remains largely unexplored. By taking a step back, this approach simplifies the theory and clarifies phenomena that are difficult to extract from more complex models. Motivated by recent advances on EPC effects in altermagnets, we address how \textit{slow phonons}, together with intrinsic spin polarization, influence the spin Edelstein response in $d$-wave altermagnets. To this end, we develop a minimal low-temperature continuum model incorporating RSOC and linear EPC (between lattice displacements and both spin-degenerate and spin-split states) at the static-Holstein level. Using the Kubo formalism, we compute the induced spin Edelstein polarization across a wide parameter range, uncovering clear signatures of the interplay between slow phonons and crystalline symmetry. 
		
		Our analysis uncovers several key insights: (i) EPC induces a net magnetization, or equivalently an out-of-plane spin polarization, and renormalizes the intrinsic spin splitting in $d$-wave altermagnets; (ii) intraband and interband transitions strongly suppress the Edelstein response with increasing EPC, eventually leading to a depolarization, where the induced spin polarization vanishes permanently; (iii) the presence of altermagnetism enhances anisotropy and modifies the usual antisymmetry between spin susceptibilities, producing directionally dependent (de)polarization; and (iv) Rashba strength and doping provide tunable knobs that shift the onset of depolarization and govern whether spin polarization is stabilized. We emphasize that phonon-induced effects are equilibrium band-structure phenomena, independent of the Edelstein effect, which is a nonequilibrium transport response. Our goal is not to invoke an additional microscopic depolarization mechanism beyond this band-structure renormalization, but to highlight that the Edelstein effect serves as a sensitive probe of the phonon-renormalized Fermi surface.
		
		\redd{We note that a static lattice
			distortion, stemming from a piezomagnetically-active strain pattern, induces a uniform spin polarization (magnetization)
			by breaking pre-existing $C_4 \mathcal{T}$ symmetry~(a combined action of time-reversal symmetry and the lattice rotation or mirror symmetry)~\cite{PhysRevMaterials.8.L041402,doi:10.7566/JPSJ.94.083702,PhysRevB.110.144421,10.1063/5.0277631,bell2026orbitalpiezomagneticpolarizabilitypure,10.1063/5.0242426,doi:10.7566/JPSJ.94.063704,doi:10.7566/JPSJ.94.083702,khodas2025tuningaltermagnetismstrain})}, allowing lattice distortions
			to convert staggered spin textures into a uniform spin imbalance.
		
		Within a fixed-chemical-potential ensemble, the Edelstein susceptibility vanishes when distortion-induced band renormalization removes all electronic states from the Fermi level. The disappearance of the response therefore reflects the collapse of
			the Fermi surface rather than a breakdown of spin-momentum locking at
			fixed carrier density. From this perspective, the Edelstein effect should be viewed as a probe of the underlying Fermi-surface topology. Our findings thus establish phonon engineering as a promising route to optimize next-generation spintronic devices. Moreover, on-demand switching between spin-polarized and depolarized states—crucial for spin logic architectures where spin rather than charge encodes information \cite{BehinAein2010}—relies on controlled depolarization to reset logic elements, erase stored information, and prevent spin leakage. 
		
		The remainder of the paper is structured as follows. Section~\ref{s2} introduces the model Hamiltonian, including the treatment of RSOC and EPC. In Sec.~\ref{s3}, we outline the calculation of spin Edelstein polarizations. Section~\ref{s4} presents the results. Brief notes on the experimental feasibility of our results are given in Sec.~\ref{s5}. Finally, Sec.~\ref{s6} summarizes key findings.\begin{figure}[t]
			\centering
			\includegraphics[width=1\linewidth]{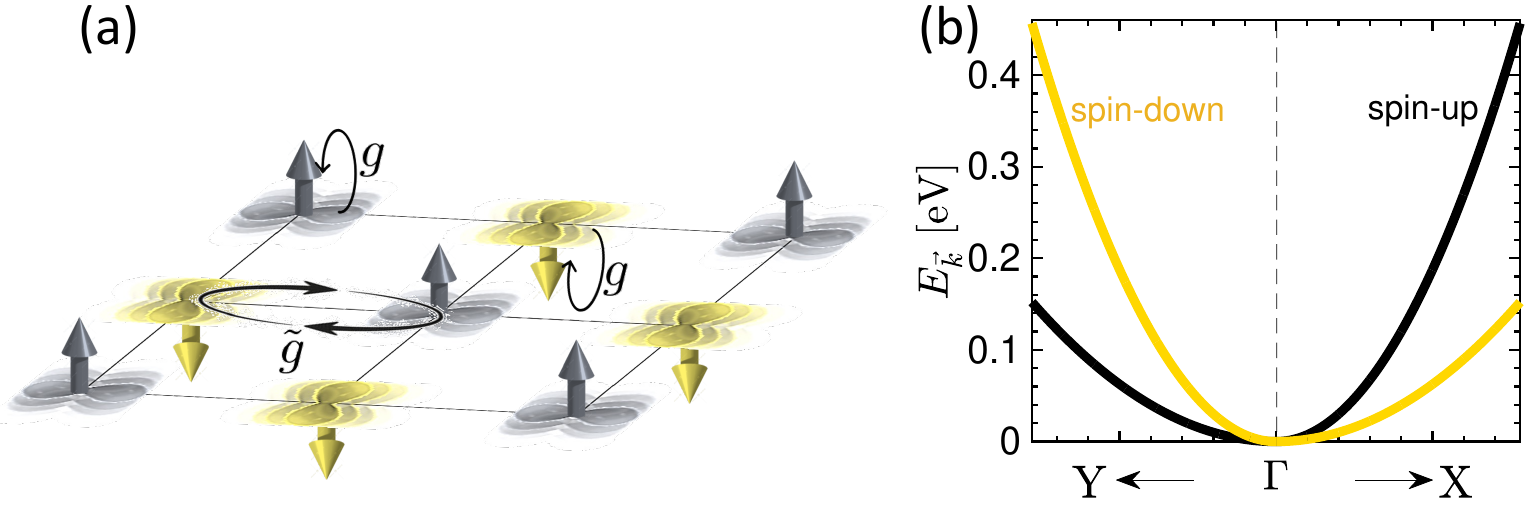}
			\caption{(a) Real-space representation of the staggered spin configuration on a square lattice of a pristine \(d_{x^2-y^2}\)-wave altermagnet with phonons included and without gating, where the spin-dependent orbital texture alternates between sublattices. The out-of-plane arrows denote two symmetry-related sublattices, $A$ and $B$, which host electrons with opposite spin polarization. Sublattice states are linearly coupled to slow Holstein phonons, depicted as shaded (vibrating) sites. Two types of EPC are shown: the conventional symmetric coupling $g$ for each sublattice/spin and the staggered antisymmetric coupling $\tilde{g}$ arising from \redd{$C_4 \mathcal{T}$ symmetry breaking}. (b) Corresponding band structure along high-symmetry paths near the $\Gamma$-point, displaying spin-split dispersions for spin-up and spin-down channels.}
			\label{f1}
		\end{figure} 
		
		\section{Hamiltonian model}\label{s2}
		\subsection{Pristine $d_{x^2-y^2}$-wave altermagnet}
		The low-energy quasiparticle excitations of a 2D Rashba-coupled $d_{x^2-y^2}$-wave altermagnet can be described by an effective two-band Hamiltonian acting in spin space. It consists of three main contributions: (i) a kinetic energy term, proportional to the identity matrix $\sigma_0$, which is spin-independent; (ii) an anisotropic spin-splitting term, proportional to $\sigma_z$, that originates from the $d$-wave altermagnetic order; and (iii) a RSOC term, proportional to $\sigma_x$ and $\sigma_y$, which breaks inversion symmetry, stemming from the gate electrodes~\cite{yarmohammadi2025anisotropiclighttailoredrkkyinteraction,PhysRevB.110.054427,cpl-42-1-017201}. The arrows in Fig.~\ref{f1}(a) denote two symmetry-related sublattices, labeled $A$ and $B$, which host electrons with opposite out-of-plane spin polarizations. Explicitly, the pristine Rashba Hamiltonian reads\begin{align}\label{eq_1}
			\mathcal{H}_{\vec{k}} = \alpha_{\vec{k}}\, \sigma_0 
			+ \beta_{\vec{k}} \, \sigma_z  
			+ \lambda_{\rm R} \left(k_y \sigma_x - k_x \sigma_y\right)\,,
		\end{align}	where $\lambda_{\rm R}$ denotes the RSOC strength. The momentum-dependent coefficients are given by\begin{align}
			\alpha_{\vec{k}} = \frac{\hbar^2}{2m_{\rm e}}\,(k_x^2 + k_y^2)\quad {\rm and} \quad
			\beta_{\vec{k}} = \frac{\hbar^2\beta}{2m_{\rm e}}\,(k_x^2 - k_y^2)\,,
		\end{align}with $m_{\rm e}$ the effective electron mass and $0 < \beta < 1$ parametrizing the $d_{x^2-y^2}$-wave anisotropy of the spin splitting. In this model, $\beta_{\vec{k}}$ even in $\vec{k}$~($-\pi/a< k_x < \pi/a$ and $-\pi/a< k_y < \pi/a$ with lattice constant $a$) preserves $C_4$ rotational symmetry, while RSOC term breaks inversion symmetry, thus, spin locks to momentum. Accordingly, RSOC wants to twist spin around the Fermi surface, while altermagnetism distorts it anisotropically. We briefly note that although our present study focuses on the 
		\(d_{x^2-y^2}\)-wave anisotropy of the spin splitting, other symmetry patterns 
		can be readily incorporated by rotating the plane of the 
		altermagnet. For instance, a \(45^\circ\) rotation of the plane yields a 
		\(d_{xy}\)-wave altermagnet characterized by \(\beta_{\vec{k}} = \frac{\hbar^2 \beta}{m_{\rm e}}\,k_x k_y\).
		
		Diagonalization of Eq.~\eqref{eq_1} yields the bare band dispersions
		\begin{equation}\label{eq_2}
			E_{\vec{k}, s} 
			= \alpha_{\vec{k}} + s \sqrt{\beta_{\vec{k}}^2 + \lambda_{\rm R}^2 k^2} 
			\equiv \alpha_{\vec{k}} +s\, d_{\vec{k}},
		\end{equation}
		where $s = \pm$ labels the spin-resolved bands, $k = \sqrt{k_x^2 + k_y^2}$, and $d_{\vec{k}}$ defines the magnitude of the effective pseudospin field. These two branches correspond to states in which the quasiparticle spin is either aligned ($+$) or anti-aligned ($-$) with the pseudospin texture generated by the combined $d$-wave anisotropy and Rashba coupling. The pristine low-energy band dispersion of \( d_{x^2 - y^2} \)-wave altermagnet with $\lambda_{\rm R} = 0$ is presented in Fig.~\ref{f1}(b). The momentum-dependent spin splitting is clearly visible. The eigenstates of Eq.~\eqref{eq_1} take the form of normalized spinors\begin{subequations}\label{eq_4}
			\begin{align}
				|\psi_{\vec{k},+}\rangle &=
				\frac{1}{\sqrt{2d_{\vec{k}}(d_{\vec{k}} + \beta_{\vec{k}})}}
				\begin{pmatrix}
					\beta_{\vec{k}} + d_{\vec{k}} \\\\
					\lambda_{\rm R}(k_y - i k_x)
				\end{pmatrix}, \\
				|\psi_{\vec{k},-}\rangle &=
				\frac{1}{\sqrt{2d_{\vec{k}}(d_{\vec{k}} + \beta_{\vec{k}})}}
				\begin{pmatrix}
					\lambda_{\rm R}(k_y + i k_x) \\\\
					-\beta_{\vec{k}} - d_{\vec{k}}
				\end{pmatrix}.
			\end{align}
		\end{subequations}
		
		\redd{The electronic Hamiltonian of an altermagnet is invariant under the combined symmetry $C_4\mathcal{T}$. Consequently, the band energies must satisfy $E_{\vec{k}}(\sigma) = E_{(C_4\mathcal{T})\, \vec{k}}\!\left((C_4\mathcal{T})\,\sigma\right)$. At the $\Gamma$ point, $\vec{k}=0$ is invariant under $C_4$ rotations, and time reversal $\mathcal{T}$ also leaves $\vec{k}=0$ unchanged. However, $\mathcal{T}$ flips the spin, $\sigma \rightarrow -\sigma$. Therefore, applying $C_4\mathcal{T}$ at $\Gamma$ yields $E_0(\sigma) = E_0(-\sigma)$, which guarantees spin degeneracy at the $\Gamma$ point. Thus, as long as $C_4\mathcal{T}$ symmetry is preserved, no spin splitting can occur at $\vec{k}=0$.}
			
		In a 2D altermagnet, spin polarization is locked to the sublattice degree of freedom rather than producing a net magnetization. The unit cell consists of two symmetry-related sublattices, $A$ and $B$, which host electrons with opposite spin polarization. Specifically, sublattice $A$ predominantly supports spin-up electrons, while sublattice $B$ supports spin-down electrons, such that the local spin expectation values satisfy $\langle \vec S_A \rangle = - \langle \vec S_B \rangle$ and the total magnetization vanishes. Despite the absence of macroscopic magnetization, the lack of combined $\mathcal{PT}$ symmetry allows for momentum-dependent spin splitting of electronic bands even in the absence of spin-orbit coupling. As a result, the Bloch states exhibit a sublattice-resolved spin texture, leading to characteristic altermagnetic spin-momentum locking protected by crystal symmetries rather than relativistic effects.
		
		\subsection{Phonon-dressed $d_{x^2-y^2}$-wave altermagnet}\label{s2b}
		We now introduce a local, spinful Holstein-type electron-phonon interaction~\cite{HOLSTEIN1959325,HOLSTEIN1959343}, see Fig.~\ref{f1}(a), described by the phonon Hamiltonian
		\begin{align}\mathcal{H}_{\text{p}} = \sum_i \left( \frac{P_i^2}{2M} + \frac{1}{2} M \omega_{\rm p}^2 Q_i^2 \right)\,,\end{align}where \(Q_i\) represents the displacement of the ion within the unit cell \(i\), \(P_i\) is the conjugate momentum, \(M\) is the ion mass, and \(\omega_{\rm p}\) is the characteristic phonon frequency. 
		
		The general linear EPC at the lattice level reads
			\begin{equation}
				H_{\mathrm{e\text{-}p}}
				= \sum_{i,\alpha,\mu} 
				\psi_i^\dagger g_\alpha^\mu \psi_i \, d_{i\alpha}^\mu ,
			\end{equation}
			where $\alpha$ labels atoms within the unit cell, $d_{i\alpha}^\mu$ atomic displacements, and $g_\alpha^\mu$ is a single-particle operator acting in the orbital/sublattice space, describing how electrons couple to a displacement (or force) of atom $\alpha$ along direction $\mu$. Expanding in scalar phonon normal modes $Q^\nu$ with polarization vectors $e_\alpha^{(\nu)\mu}$ gives $	d_{i\alpha}^\mu = \sum_\nu e_\alpha^{(\nu)\mu} Q^\nu_i$ and
			$H_{\mathrm{e\text{-}p}}
				= \sum_{i,\nu} Q^\nu_i
				\psi_i^\dagger 
				\Big[\sum_{\alpha,\mu} g_\alpha^\mu e_\alpha^{(\nu)\mu}\Big] 
				\psi_i .$ Projecting onto a $s$-band subspace, $\psi_i = \sum_s c_i^s|u_s\rangle$, where $c^s_i$ is the fermionic annihilation operator for an electron in band $s$ localized at unit cell $i$, encoding the amplitude/operator degree of freedom associated with all atomic orbitals within unit cell $i$ that contribute to band $s$, we define
			\begin{equation}
				(\hat{O}_\nu)_{s's} \equiv \sum_{\alpha,\mu} \langle u_{s'}|g_\alpha^\mu|u_s\rangle \, e_\alpha^{(\nu)\mu},
			\end{equation}so that the Hamiltonian becomes
			$H_{\mathrm{e\text{-}p}} = \sum_{i,\nu} Q^\nu_i\, c^\dagger_i \hat{O}_\nu c_i$. The matrix element $\langle u_{s'} |
			g_{\alpha}^{\mu}
			| u_s \rangle$ is the band-resolved coupling vertex associated with moving atom $\alpha$ along direction $\mu$, giving rise to the corresponding intra- and interband coupling amplitudes. All directional and atomic information enters only parametrically via $\hat{O}_\nu$. Different lattice displacements therefore correspond simply to different values of these coefficients.
		 	
		 With adiabatic phonons ($\nu=0$) and following the spinors in Eq.~\eqref{eq_4} for our low-energy model in the presence of static lattice distortion, as well as Eq.~\eqref{eq_1} with $\lambda_{\rm R} = 0$, $\hat{O}_0$ decomposes as~(see Appendix~\ref{ap1} for a detailed derivation)
		 \begin{equation}\label{eq_10}
		 	\hat{O}_0 =\frac{g_A^x + g_B^y}{2}\sigma_0 + \frac{g_A^x -g_B^y}{2}\sigma_z \, .
		 \end{equation}\redd{This is a minimal model as a proof-of-principle, demonstrating how the pristine Rashba altermagnet can be modified by a static lattice distortion. The $\sigma_0$ term corresponds to a sublattice-symmetric deformation potential, while the $\sigma_z$ term represents a staggered EPC that breaks sublattice symmetry and can generate a Zeeman-like contribution in the altermagnetic phase. In the pristine altermagnet, the degeneracy at the $\Gamma$ point is protected by the combined action of time-reversal symmetry and the lattice rotation (or mirror) symmetry that exchanges the two sublattices, i.e., by the $C_4 \mathcal{T}$ symmetry. To have the $\sigma_z$ term, we propose a static lattice distortion to break $C_4 \mathcal{T}$ symmetry. This distortion drives atomic displacements within the unit cell, thereby lowering the symmetry. This can occur in three distinct ways: by breaking the $C_4$ rotational symmetry, by breaking time-reversal ($\mathcal{T}$) symmetry, or by breaking both $C_4$ and $\mathcal{T}$ such that their combined operation is no longer a symmetry. In all cases, the protecting $C_4 \mathcal{T}$ symmetry is no longer present and an an out-of-plane Zeeman-like term of the form $\propto \sigma_z$ can be generated. To only break $C_4$, one can think of structural anisotropy that lowers rotation symmetry and this is physically possible through uniaxial strain, distortion, or substrate-induced anisotropy~\cite{PhysRevMaterials.8.L041402,doi:10.7566/JPSJ.94.083702,PhysRevB.110.144421,10.1063/5.0277631,bell2026orbitalpiezomagneticpolarizabilitypure,10.1063/5.0242426,doi:10.7566/JPSJ.94.063704,doi:10.7566/JPSJ.94.083702,khodas2025tuningaltermagnetismstrain}. To only break $\mathcal{T}$, one can add a Zeeman magnetic field to destroy the altermagnetic protection~\cite{PhysRevB.110.024425}. To break both $C_4$ and $\mathcal{T}$, one can apply a circularly polarized light (Floquet), which breaks $\mathcal{T}$ and mirror symmetries~\cite{xt23-9pnv,k3xb-8pts,yarmohammadi2026efficienttwocolorfloquetcontrol}.}
		 
		 \redd{In practice, a physically realistic mechanism must involve a specific symmetry-lowering distortion pattern that explicitly breaks the relevant $C_4 \mathcal{T}$ symmetry. Since we involve a deformation process by considering phonons, breaking $C_4$ is the best solution. We identify a piezomagnetically active strain~\cite{PhysRevMaterials.8.L041402,doi:10.7566/JPSJ.94.083702,PhysRevB.110.144421,10.1063/5.0277631,bell2026orbitalpiezomagneticpolarizabilitypure,10.1063/5.0242426,doi:10.7566/JPSJ.94.063704,doi:10.7566/JPSJ.94.083702,khodas2025tuningaltermagnetismstrain} as a plausible microscopic origin of the static lattice distortion considered here. This is both theoretically well-justified and experimentally relevant. However, a detailed analysis of the underlying symmetry-breaking mechanism—such as that discussed in Ref.~\cite{khodas2025tuningaltermagnetismstrain}—goes beyond the scope of the present minimal model and will be addressed in future work.}

By setting $Q_i^0 = Q_i$, the distortion-induced electron–phonon Hamiltonian can be written as
		\begin{align}\label{eq_6n}
			\mathcal{H}_{\text{e-p}} =  \sum_i \left[g^x_A \left(n_{i,A} -\tfrac{1}{2}\right) + g^y_B \left(n_{i,B} -\tfrac{1}{2}\right)\right] Q_i\,.
		\end{align}The shaded sites in Fig.~\ref{f1}(a) schematically represent a staggered phonon mode that couples to the electronic density. The subtraction of $1/2$ from the electron number ensures that the phonon coordinate $Q_i$ couples only to deviations from half-filling, so that $Q_i = 0$ corresponds to the undistorted equilibrium lattice~\cite{PhysRevB.61.R838,PhysRevX.14.031052,PhysRevE.105.025301,PhysRevB.60.7950,faundez2024twodimensionalrashbaholsteinmodel}. Thus, the EPC is nonzero only when the electron density deviates from $1/2$, leading to a finite phonon displacement. Since $g_A^x$ acts only on sublattice $A$ with spin-up and $g_B^y$ only on
		sublattice $B$ with spin-down, we choose $g_A^x = g_{\uparrow}$ and $g_B^y = g_{\downarrow}$ across the system. On the other hand, since we do not perform \emph{ab initio} calculations to determine direction-dependent EPC constants for different displacement patterns, we introduce
		the symmetric and antisymmetric combinations (see Fig.~\ref{f1}(a))\begin{align}
			g = \frac{g_{\uparrow} + g_{\downarrow}}{2}, \quad \tilde{g} = \frac{g_{\uparrow} - g_{\downarrow}}{2}\, ,
		\end{align}in which we define the spin-dependent couplings as \(g_{\uparrow} = g + \tilde{g}\) and \(g_{\downarrow} = g - \tilde{g}\). In the absence of precise values for \( g_\uparrow^x \) and \( g_\downarrow^y \),
		sweeping a broad parameter range for \( g \) and \( \tilde{g} \) effectively spans
		the relevant lattice-displacement directions.
		
		Substituting these into the general EPC term in Eq.~\eqref{eq_6n}, we obtain\begin{align}
			\mathcal{H}_{\text{e-p}} 
			=  \sum_i \left[g \bigl(n_{i,\uparrow} + n_{i,\downarrow}-1\bigr) 
			+ \tilde{g}\bigl(n_{i,\uparrow} - n_{i,\downarrow}\bigr) \right] Q_i \,.
		\end{align}
		This expression naturally separates into two contributions:  
		a symmetric part coupling to the total local electron density \(n_i = n_{i,\uparrow} + n_{i,\downarrow}\)  
		and an antisymmetric part coupling to the \(m_{{\rm s},i} = n_{i,\uparrow} - n_{i,\downarrow}\), the difference between electronic densities on the two altermagnetic sublattices. In our model, the bilinear term $Q\,(n_\uparrow - n_\downarrow)$ does not represent a local spin on a single site, but the staggered, sublattice-resolved electron density in the altermagnet. \redd{Since the symmetry-lowering piezomagnetically-active strain patterns break $C_4 \mathcal{T}$ symmetry~\cite{PhysRevMaterials.8.L041402,doi:10.7566/JPSJ.94.083702,PhysRevB.110.144421,10.1063/5.0277631,bell2026orbitalpiezomagneticpolarizabilitypure,10.1063/5.0242426,doi:10.7566/JPSJ.94.063704,doi:10.7566/JPSJ.94.083702,khodas2025tuningaltermagnetismstrain}, this coupling is allowed}. Therefore, no compensating frequency is required, and the coupling is fully consistent with the symmetry of the altermagnetic Hamiltonian. 
		
		Assuming a staggered phonon displacement pattern across the sublattices but with a static-Holstein displacement amplitude, we focus on the regime of slow phonons where the lattice responds quasi-statically to the electronic environment, i.e., we set \(\langle Q_i \rangle = Q_0\). The slow phonon-dressed Hamiltonian in momentum space can be written within the static-Holstein approximation as{\small \begin{align}\label{eq_6}
				\mathcal{H}_{\vec{k}}^{\text{SH}} = \left( \alpha_{\vec{k}} + g Q_0 \right) \sigma_0 + \left(\beta_{\vec{k}}+ \tilde{g} Q_0\right) \sigma_z + \lambda_{\rm R}  \left(k_y \sigma_x - k_x \sigma_y\right).
		\end{align}}The phonon-induced terms \(g Q_0\) and \(\tilde{g} Q_0\) play distinct roles in the low-energy electronic structure. The shift \(g Q_0\) renormalizes the chemical potential and provides a route for phonon-assisted tuning of the carrier density. In contrast, the spin-dependent term \(\tilde{g} Q_0\) acts as a Zeeman-like field, lifting the spin degeneracy and enabling transient spin polarization or spin-current control. When \(\tilde{g}=0\), the spectrum therefore remains isotropic along the \(\Gamma\! \to\! X\) and \(\Gamma\! \to\! Y\) directions. By contrast, a finite \(\tilde{g}\) introduces opposite shifts for spin-up and spin-down bands, leading to spin-dependent anisotropies even though the coupling itself is momentum independent.
	
		\redd{While the present model is formulated at a minimal, proof-of-principle level, the breaking of $C_4 \mathcal{T}$ symmetry induced by a static lattice distortion remains operative in realistic multi-sublattice systems~\cite{doi:10.7566/JPSJ.88.123702,PhysRevB.106.094432}. In particular, if a staggered field emerges from the interplay between magnetic and nonmagnetic sublattices  mediated by lattice distortions in realistic materials, this still validates the underlying symmetry-based mechanism discussed here. Although material-specific details may renormalize the effective electronic-structure parameters and the strength of the EPCs, the fundamental distortion-induced phenomena described here remain symmetry-allowed and robust.}
		
		The equilibrium phonon displacement \(Q_0\) is determined by minimizing the total free energy, taking into account both the harmonic restoring force of the lattice and the coupling to the electron density:\begin{align}
			M \omega_{\rm p}^2 Q_0 = {} - g\, [n_{\rm e}-1] - \tilde{g}\, m_{\rm s}\, ,
		\end{align}where $n_{\rm e} = n_\uparrow + n_\downarrow$ is the total density and $m_{\rm s} = n_\uparrow - n_\downarrow$ is the imbalance density, with
		\begin{subequations}
			\begin{align}
				n_{\rm e} = {} &\frac{1}{N_{\vec k}} \sum_{\vec{k},s} f_{\vec{k},s} \langle \tilde{\psi}_{\vec{k},s} | \sigma_0 | \tilde{\psi}_{\vec{k},s} \rangle \,,\\
				m_{\rm s} = {} &\frac{1}{N_{\vec k}} \sum_{\vec{k},s} f_{\vec{k},s} \langle \tilde{\psi}_{\vec{k},s} | \sigma_z | \tilde{\psi}_{\vec{k},s} \rangle \,,
			\end{align}
		\end{subequations}where $\tilde{\psi}$ is phonon-dressed spinor, \(N_{\vec k}\) denotes the number of \(\vec{k}\)-points in the Brillouin zone, and \(f_{\vec{k},s}\) is the Fermi-Dirac distribution function, $f_{\vec{k},s} = (e^{(\mathcal{E}_{\vec{k},s} - \mu) / k_{\rm B} T} + 1)^{-1}$; $\mu$ is the bare chemical potential and $\mathcal{E}_{\vec{k},s}$ is the phonon-dressed band dispersion:\begin{align}\label{eq_12o}
			\mathcal{E}_{\vec{k}, s} 
			= \alpha_{\vec{k}} + g Q_0 +s \sqrt{\left(\beta_{\vec{k}} + \tilde{g} Q_0\right)^2 + \lambda_{\rm R}^2 k^2}	\equiv \tilde{\alpha}_{\vec{k}} +s\, \tilde{d}_{\vec{k}}.
		\end{align}The phonon-dressed quasiparticle dispersion inherits several key features from the underlying electronic structure and the EPC. The first term,
		\(\tilde{\alpha}_{\vec{k}} = \alpha_{\vec{k}} + g Q_0\),
		represents a uniform shift of all bands due to the static lattice distortion \(Q_0\).
		The second term,
		\(\tilde{d}_{\vec{k}} = \sqrt{(\beta_{\vec{k}} + \tilde{g} Q_0)^2 + \lambda_{\rm R}^2 k^2}\),
		controls the spin splitting between \(s = \pm 1\) bands.
		Importantly, \(\beta_{\vec{k}} + \tilde{g} Q_0\) is an even but anisotropic function of momentum:
		the coupling \(\tilde{g} Q_0\) modifies its magnitude in a direction-dependent way,
		preserving inversion symmetry but breaking \(C_4\) rotational symmetry.
		As a result, the dressed bands show spin-dependent gaps and anisotropic distortions whose strength grows with \(Q_0\),
		explaining the deformation.
		The Rashba term \(\lambda_{\rm R}^2 k^2\) further broadens the spin splitting,
		ensuring that even when \(\beta_{\vec{k}} + \tilde{g} Q_0 \to 0\) along some directions,
		the bands remain separated by a gap proportional to \(|\lambda_{\rm R} k|\).
		
		Since our effective model is valid only within a restricted energy window, we focus on very low temperatures, ensuring that thermal excitations remain within the model’s regime of validity. In the zero-temperature limit, the Fermi-Dirac occupations reduce to $f_{\vec{k},s} \approx \Theta(\mu - \mathcal{E}_{\vec{k},s})$, where \(\Theta\) is the Heaviside function. Therefore, $n_{\rm e}$ and $m_{\rm s}$ satisfy the self-consistent equations\begin{subequations}\label{eq_9}\begin{align}
				n_{\rm e} \approx {} &\frac{1}{N_{\vec k}} \sum_{\vec{k}} \left[\Theta\left(\mu  - \mathcal{E}_{\vec{k},+}(Q_0)\right) + \Theta\left(\mu- \mathcal{E}_{\vec{k},-}(Q_0)\right)\right] \,,\label{eq_9a}\\
				m_{\rm s} \approx {} &\frac{1}{N_{\vec k}} \sum_{\vec{k}} \left[\Theta\left(\mu- \mathcal{E}_{\vec{k},+}(Q_0)\right) - \Theta\left(\mu- \mathcal{E}_{\vec{k},-}(Q_0)\right)\right]\,,\label{eq_9b}
		\end{align}\end{subequations}where \(Q_0\) depends on both \(n_{\rm e}\) and $m_{\rm s}$. Within the static-Holstein approximation, $Q_0$
		renormalizes the band energies according to Eq.~\eqref{eq_12o}, independent of
		any external electric field. As the EPC increases, the phonon-induced energy shifts
		can push the spin-split bands entirely above (or below) the chemical
		potential, leading to a complete disappearance of Fermi-surface states.
		We refer to this phenomenon as a phonon-induced Fermi-surface collapse.
		This collapse is an equilibrium effect determined solely by the band
		renormalization and does not rely on transport or response functions. In the following sections, we show that the vanishing of the Edelstein
		spin polarization occurs only after this Fermi-surface collapse,
		demonstrating that the depolarization is a consequence—not the origin—of
		the equilibrium band reconstruction.
		
		\section{Spin Edelstein effect}\label{s3}
		In noncentrosymmetric crystals, spin-orbit interaction enables an external electric field to induce a nonequilibrium spin or orbital polarization. When involving electron spin, this phenomenon is known as the spin Edelstein (Rashba-Edelstein) effect~\cite{edelstein1990spin,Salemi2019,Johansson_2024}. It is crucial to distinguish between \emph{spin splitting} and \emph{spin polarization}. Spin splitting refers to the intrinsic energy separation of bands with opposite spin projections, originating from altermagnetic order and/or RSOC, and is encoded in the equilibrium band structure even in the absence of external perturbations. In contrast, spin polarization denotes an induced magnetization arising from nonequilibrium population imbalance between spin states, generated here by electron-phonon interactions or electric fields. Unlike spin splitting, spin polarization is not an intrinsic property of the electronic structure but is strongly tunable by lattice distortions and external driving.
		
		\redd{While Fermi-surface modulations due to the EPC associated with possible metal-to-(band)insulator transitions can be detected using thermodynamic quantities such as the heat capacity, most of these probes are sensitive only to the total density of states. Here, our focus is on how the spin Edelstein response evolves as the Fermi surface shrinks. Unlike conventional thermodynamic measures, the Edelstein tensor provides symmetry-resolved information about the spin structure of the electronic states, making it a particularly sensitive probe of distortion-controlled altermagnetic behavior.}
		
		Within the linear response regime, the induced expectation value of a spin operator \(\hat{S}^\ell\) (\(\ell \in \{x, y, z\}\)) due to an in-plane electric field \(\mathbb{E}^j\) along direction \(j \in \{x, y\}\) can be written as~\cite{PhysRevMaterials.5.074407,PhysRevMaterials.6.095001}\begin{align}
			\langle \hat{S}^\ell \rangle = \sum_j \chi_{\ell j} \mathbb{E}^j \,,
		\end{align}where \(\chi_{\ell j}\) defines the linear susceptibility, also called the Edelstein response tensor. It contains contributions from intraband processes within a single band as well as from interband coherences, given by\begin{align}\label{eq_12}
			\chi_{\ell j} = {} &\frac{e}{4 \pi^2} \int d^2k \Bigg[ \tau_{\rm intra}\,
			\sum_{s} \frac{\partial f_{\vec{k},s}}{\partial \mathcal{E}_{\vec{k}, s} } S^\ell_{\vec{k},ss}\, v^j_{\vec{k},ss} 
			\notag \\ {} &	- i\sum_{s \neq s'} \frac{ f_{\vec{k},s}-  f_{\vec{k},s'}}{\mathcal{E}_{\vec{k}, s}  - \mathcal{E}_{\vec{k}, s'} } 
			\frac{S^\ell_{\vec{k},s' s}\, v^j_{\vec{k},s s'}}{\mathcal{E}_{\vec{k}, s'}  - \mathcal{E}_{\vec{k}, s} + i / \tau_{\rm inter}} \Bigg] \,.
		\end{align}In this study, we set the intraband and interband broadening times to be equal, $\tau^{-1}_{\rm intra} = \tau^{-1}_{\rm inter} = 0.5~\mathrm{eV}$, a value consistent with typical metallic systems~\cite{PhysRevMaterials.6.095001,Salemi2019}. We note that all energies appearing in Eq.~\eqref{eq_12} correspond to the phonon-dressed quasiparticle energies defined in Eq.~\eqref{eq_12o}. The operator matrix elements are defined as\begin{subequations}
			\begin{align}
				S^\ell_{\vec{k},s's} = {} &\frac{\hbar}{2}\langle\tilde{\psi}_{\vec{k},s'} | \hat{\sigma}^\ell | \tilde{\psi}_{\vec{k},s} \rangle\, ,\\
				v^j_{\vec{k},s s'} = {} &\frac{1}{\hbar}\langle \tilde{\psi}_{\vec{k},s} | (\partial \mathcal{H}_{\vec{k}}^{\text{SH}} / \partial k_j) | \tilde{\psi}_{\vec{k},s'} \rangle\, .
		\end{align}	\end{subequations}
	
	Our evaluation of the Edelstein response employs a constant relaxation-time approximation. \redd{In the present static-Holstein framework, the ladder-type vertex corrections vanish within linear response~\cite{PhysRevB.109.245132,PhysRevB.109.214312}. This is because the phonon coordinate is replaced by its equilibrium value $Q_0$, yielding a purely static and quadratic Hamiltonian, with the phonon effects entering only as renormalizations of the single-particle. Thus, the resulting theory does not involve a dynamical phonon propagator, frequency-dependent electron self-energy, or phonon-induced scattering. In this strictly static framework, therefore, there are no phonon-induced vertex corrections in the many-body sense. A full treatment of vertex corrections in the presence of dynamical phonons is beyond our current scope and will be left for future investigation.}\begin{figure*}[t]
		\centering
		\includegraphics[width=1\linewidth]{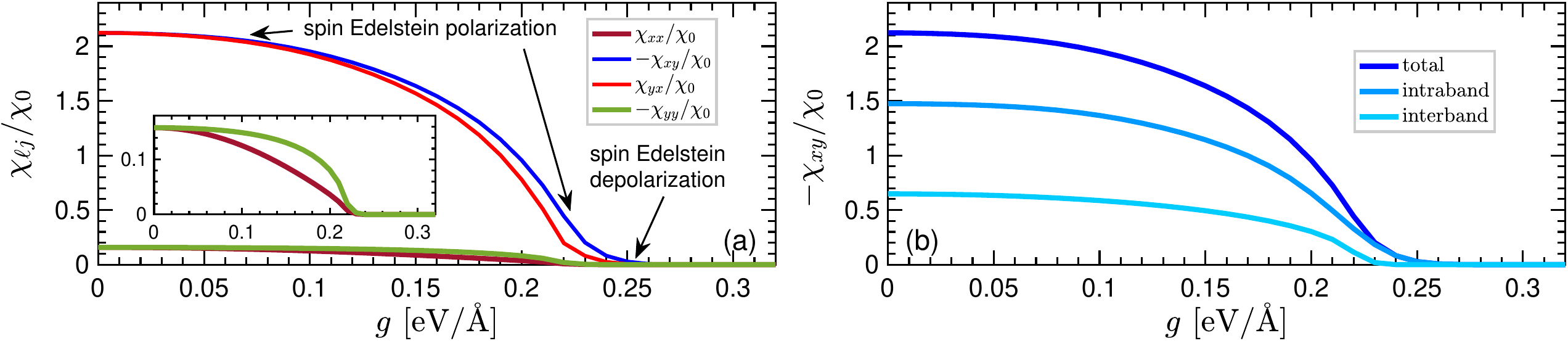}
		\caption{(a) Components of the spin Edelstein susceptibility tensor, $\chi_{\ell j}/\chi_{0}$, as a function of the EPC strength \( g \) for fixed altermagnetic strength 
			\(\beta = 0.5\), RSOC strength \(\lambda_{\rm R} = 0.3\) eV$\cdot$\AA, and chemical potential \(\mu = 0\) at Fermi energy. The susceptibilities start at a nearly constant value 
			for weak coupling, reflecting robustness of the spin response 
			in the perturbative regime, and then decrease monotonically with 
			increasing $g$, eventually approaching zero at moderate-to-strong 
			coupling, marking the onset of spin Edelstein depolarization. (b) Decomposition of \(-\chi_{xy}/\chi_{0}\) into its intraband and 
			interband contributions, along with its total response. The intraband processes provide an increasingly significant contribution, and together with the interband effects, they ultimately lead to the complete suppression of the spin Edelstein polarization at $g_{\rm c} \approx 0.25$ eV/\AA.}
		\label{f2}
	\end{figure*}
		
		With the phonon-dressed Hamiltonian from Eq.~\eqref{eq_6}, the corresponding matrix elements can be computed. In the zero-temperature limit, the derivative of the Fermi function reduces to a Dirac delta $\frac{\partial f}{\partial \mathcal{E}} \to -\delta(\mathcal{E} - \mu)$. Substituting this into the intraband term yields a simplified expression for the low-temperature response \begin{align}
				\chi^{\rm intra}_{\ell j} = - \frac{e}{4 \pi^2} \tau_{\rm intra} \int d^2k \sum_s \delta(\mathcal{E}_{\vec{k},s} - \mu)\, S^\ell_{\vec{k},ss}\, v^j_{\vec{k},ss}\,.
			\end{align}This form directly connects the intraband Edelstein response to the states at the Fermi surface and their corresponding operator and velocity matrix elements.	
		
		In the following, we compute the spin Edelstein effect within our phonon-dressed model, which accounts for the slow phononic effects. This allows us to investigate how the distortion-induced renormalization of the chemical potential, $g Q_0$, and the generation of an effective magnetization, $\tilde{g} Q_0$, influence the spin Edelstein polarization. In particular, we analyze how these phonon-mediated modifications compete or cooperate with other intrinsic parameters of the model to shape the magnitude and orientation of the induced spin Edelstein polarization. In our simulations, for slow phonon, we take the limits \(M \to \infty\) and \(\omega_{\rm p} \to 0\) such that the product \(M \omega_{\rm p}^2\), corresponding to the stiffness of the phonon potential, is fixed to unity.
		
		Throughout this work, the limit $\omega_{\rm p} \to 0$ should be understood as
			the adiabatic (static) limit of an internal lattice distortion, rather
			than as the long-wavelength limit of an acoustic phonon. \redd{In other words, the limit $\omega_{\rm p} \to 0$ is taken only in the Born--Oppenheimer sense~\cite{Ziegler_2005}, rather than a dynamical phonon mode softening to zero frequency, ensuring a classical lattice field with fixed stiffness $M \omega_{\rm p}^2$.}
			While a uniform acoustic displacement corresponds to a rigid
			translation and does not affect the electronic spectrum, the phonon
			mode considered here represents a zone-center optical distortion with
			relative atomic displacements inside the unit cell.
			Such a distortion generates a finite local potential and therefore
			leads to a nonvanishing EPC even in the static
			limit. 
			
			\redd{We briefly note that the Edelstein effect, describing current-induced spin polarization via a shift of a spin-textured Fermi surface, remains a useful conceptual framework in our system~\cite{Chakraborty2025}. Under an applied electric field, the electronic distribution acquires a momentum shift, and the resulting spin polarization arises from the spin expectation values of the occupied Bloch states at the displaced Fermi surface. In the limit of dominant RSOC and negligible altermagnetic splitting, our model smoothly reduces to the conventional Rashba 2D electron gas, reproducing the familiar helical spin texture. In contrast, in the altermagnetic regime with strain-induced symmetry lowering, the spin texture is qualitatively modified. While we do not explicitly show the helical spin configuration in momentum space, through band structure analysis in the following, we illustrate how the momentum-dependent exchange splitting and sublattice symmetry breaking reshape the Fermi surface geometry.}
		
		\section{Results and discussion}\label{s4}
		The key model parameters are the EPCs $g$ (on-site) and $\tilde{g}$ (staggered \redd{arising from $C_4 \mathcal{T}$ symmetry breaking}), the altermagnetic strength $0<\beta<1$, the RSOC strength $\lambda_{\rm R}$, and the chemical potential $\mu$. The EPCs are constrained by the Lindemann criterion~\cite{Lindemann1910}, which limits atomic displacements to $\lesssim10\%$ of the lattice constant. For representative $d$-wave altermagnets with lattice constants of $3$--$4\,\text{\AA}$, we choose $g$ such that the equilibrium displacement satisfies $Q_0\!\le\!0.3\,\text{\AA}$. Since the staggered energy scale is smaller than the on-site hopping, we take $\tilde{g}<g$ and set $\tilde{g}=g/4$ unless stated otherwise. Finally, $\lambda_{\rm R}$ and $\mu$ are varied across the spin-channel bandwidth to ensure a proper alignment of the spin response with the electronic structure.
		
		In the numerical simulations, Eqs.~\eqref{eq_9a} and~\eqref{eq_9b} are solved self-consistently for given EPCs and bare chemical potential $\mu$~\cite{PhysRevB.68.165102}. For specified $n_{\rm e}$ and $m_{\rm s}$, the equilibrium phonon displacement is $Q_0 = -\left(g\,[n_{\rm e}-1] + \tilde{g}\,m_{\rm s}\right)/M\omega_{\rm p}^2$.
		Although $Q_0$ arises within a static-Holstein framework, it corresponds to a stable minimum of the full Hamiltonian and remains finite and self-consistent. Our approach thus constitutes a self-consistent, quasi-static treatment in which $Q_0$ is determined by minimizing the coupled electron-phonon Hamiltonian. In the slow-phonon limit ($\omega_{\rm p}\!\to\!0$), inelastic scattering and dissipative processes are neglected. While dynamical fluctuations and frequency-dependent self-energies are neglected, the method captures nonperturbative, density-driven lattice shifts beyond linearized treatments, corresponding formally to the zero-frequency limit of lowest-order Migdal theory~\cite{Allen1983}. We stress that including dynamical phonons would introduce self-energy corrections and finite lifetimes, potentially smoothing sharp spectral features, but lies beyond the present scope. 
		
		Once the complete Hamiltonian in Eq.~\eqref{eq_6} is established, we can apply the Kubo formula from Eq.~\eqref{eq_12} to calculate the spin Edelstein effect. Within this approach, the linear spin response to an applied electric field is computed from the eigenvalues and eigenstates of the phonon-dressed Hamiltonian, thereby probing phonon-induced renormalization effects.
		
		Figure~\ref{f2}(a) illustrates the evolution of the 
		spin Edelstein susceptibilities, $\chi_{\ell j}/\chi_{0}$, where $\chi_{0} = e/4\pi^2$ sets the reference scale, with increasing EPC strength $g$ at $\mu = 0$. For weak EPC (\(g \lesssim 0.06\) eV/\AA), the susceptibilities
		remains nearly constant. In this regime, electron-phonon interaction is too weak to significantly modify the spin-split band 
		structure, which reflects the perturbative regime. 
		In other words, the spin-momentum locking characteristic of Rashba altermagnet ensures a 
		robust spin Edelstein response in this regime. To explicitly demonstrate the breaking of the 
		antisymmetry relation in the spin Edelstein response, $\chi_{\ell j} \neq -\chi_{j \ell}$, we plot the quantity \(-\chi_{j \ell}\) for comparison. If the antisymmetric 
		condition \(\chi_{\ell j} = -\chi_{j \ell}\) were preserved under the influence of EPC, 
		the curves corresponding to \(\chi_{\ell j}\) and \(-\chi_{j \ell}\) would coincide, 
		exhibiting no deviation. We note that, owing to the symmetry of our model---specifically, \( v^z_{{\vec{k}},ss'} = 0 \) arising from the absence of a \( k_z \) component in the Hamiltonian---the spin Edelstein effect cannot generate out-of-plane (\( z \)-) polarization, leading to \( \chi_{\ell z} = \chi_{z j} = 0 \). Thus, only in-plane spin polarizations are induced.
		
		For all $g$ values, the dominant contributions arise from the off-diagonal elements 
		$-\chi_{xy}$ and $\chi_{yx}$. These terms 
		encode the transverse Edelstein response, reflecting the strong spin-momentum 
		locking induced by RSOC. In contrast, the diagonal 
		components $\chi_{xx}$ and $-\chi_{yy}$ remain much smaller, with $\chi_{xx}$ 
		slowly decreasing and $-\chi_{yy}$ staying nearly constant across a large 
		range of $g$. As $g$ increases, both $-\chi_{xy}$ and $\chi_{yx}$ undergo a monotonic decrease, 
		eventually touching zero near a threshold coupling $g_{\rm c} \approx 0.25$ eV/\AA. This zero response signals the point of \emph{spin Edelstein 
			depolarization}.  The coupling strength $g_{\rm c}$ defines a crossover scale at which the
			phonon-induced band renormalization removes the Fermi surface.
			No symmetry change or phase transition is associated with this point. Accordingly, the term ``transition'' is used here in a pragmatic sense to
			denote the onset of complete Fermi-surface depletion, rather than a
			sharp thermodynamic transition. \redd{It should be emphasized that the strong suppression of the Fermi surface is not the result of an unrealistically large lattice deformation. As noted before, the Lindemann criterion~\cite{Lindemann1910} constrains atomic displacements to $\lesssim 10\%$ of the lattice constant, ensuring that the static lattice distortion remains within a physically reasonable range.}\begin{figure}[t]
				\centering
				\includegraphics[width=1\linewidth]{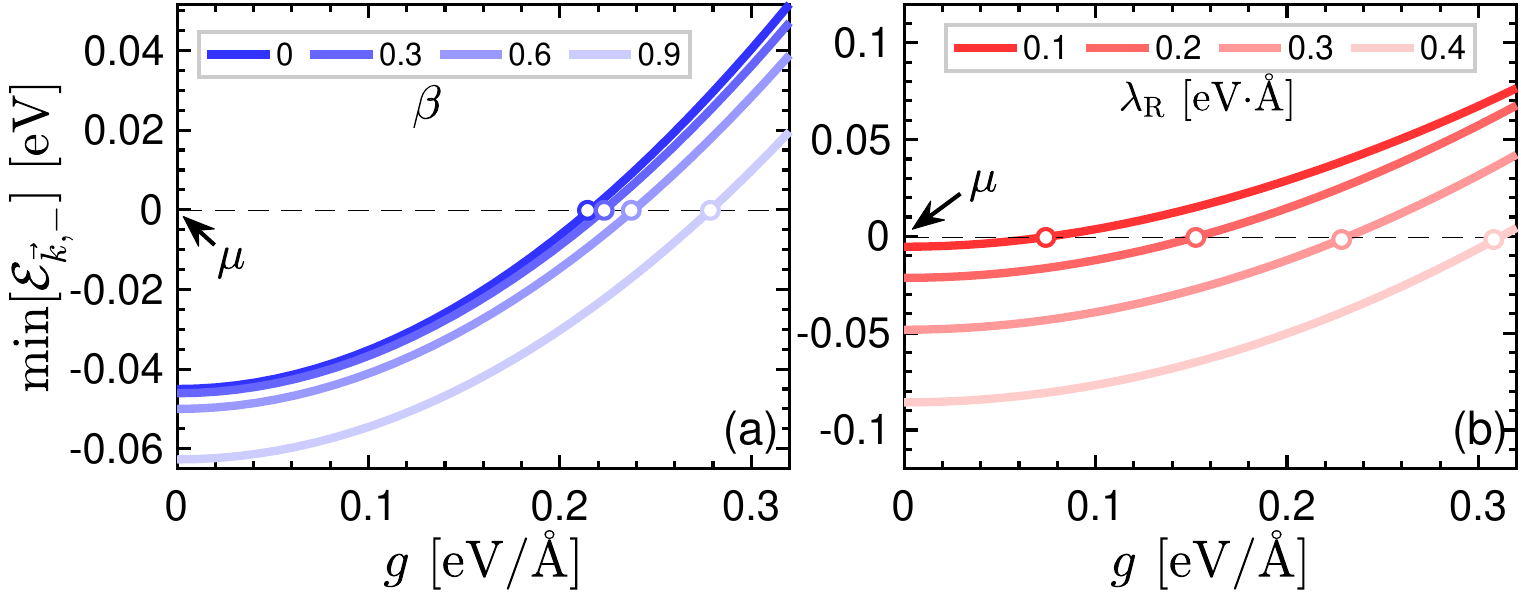}
				\caption{Minimum of the lower spin-split band, $\min[\mathcal{E}_{\vec{k},-}]$, as a function of the EPC $g$ for two parameter variations. 
					(a) Variation of the altermagnetic anisotropy $\beta$ and  
					(b) variation of the RSOC strength $\lambda_{\rm R}$. 
					The horizontal dashed line marks the zero Fermi energy $\mu$. Open circles indicate the threshold EPC $G_{\rm c}$ at which the lower band touches $\mu$, marking the onset of spin Edelstein depolarization at $g_{\rm c} 
					\gtrsim G_{\rm c}$.}
				\label{f3_new}
			\end{figure}
		
		The distinct vanishing of 
		$-\chi_{xy}$ and $\chi_{yx}$ indicates that the depolarization affects both 
		directions of the transverse spin response anisotropically and can be rationalized from symmetry considerations in the phonon-renormalized band structure. From Eq.~\eqref{eq_12o}, the anisotropic term $\beta_{\vec{k}} + \tilde{g} Q_0$ modifies the spin-split eigenvalues differently along the $\Gamma\!\to\!X$ and $\Gamma\!\to\!Y$ directions. Since $\chi_{xy}$ and $\chi_{yx}$ are sensitive to the momentum-resolved band dispersions, the anisotropic shift induced by the phonon term causes the conditions for depolarization (i.e., when certain Fermi states no longer contribute) to occur at different EPC strengths along these directions. In other words, the symmetry-breaking effect of $\tilde{g} Q_0$ on $\tilde{d}_{\vec{k}}$ lifts the degeneracy between $x$ and $y$ susceptibilities, naturally leading to distinct vanishing points for $\chi_{xy}$ and $\chi_{yx}$ (see Fig.~\ref{f3} for confirmation).
		
		Figure~\ref{f2}(b) provides further insight by decomposing the transverse component 
		$-\chi_{xy}/\chi_{0}$ into its intraband and interband contributions. A similar analysis can be carried out for the other components.  
		The intraband part dominates at all coupling strengths, reflecting the 
		usual Edelstein mechanism in which current-carrying states at the Fermi surface 
		induce spin polarization. The interband part remains in relative weight with $g$. 
		
		To understand the depolarization, we examine the spin-resolved band structure in the presence of slow static-Holstein phonons. Revisiting the spin susceptibility in Eq.~\eqref{eq_12}, vanishing $\chi_{\ell j}$ requires the velocity $v^j_{\vec{k}}$ to vanish in the dominant regions of the Brillouin zone. This occurs, for example, when the spin-split bands are shifted away from the Fermi energy due to $\mu=0$ in our simulations. In such cases, the group velocity $v^j_{\vec{k}} = \hbar^{-1}\partial \mathcal{E}_{\vec{k}}/\partial k_j$ approaches zero because there are no bands near the Fermi level. Consequently, both the charge current and the induced spin polarization vanish.
		
		To identify the threshold coupling strength \(g_{\rm c}\) at which the spin Edelstein polarization is fully suppressed, marking the onset of depolarization, we consider the band dispersion $\mathcal{E}_{\vec{k}, s}$. Since the lower band (\(s=-1\)) gives the smallest energies, the condition for the absence of zero-energy states is $
		\min_{\vec{k}}[\mathcal{E}_{\vec{k}, -}] > 0$. 
		
		Figure~\ref{f3_new} shows the evolution of the minimum of the lower spin-split band as the EPC $g$ is increased. The threshold coupling $G_{\rm c}$ is defined by $\min_{\vec{k}}[\mathcal{E}_{\vec{k},-}] = 0$ and depends on the parameter set. In Fig.~\ref{f3_new}(a), increasing the altermagnetic anisotropy $\beta$ shifts the curves downward, delaying depolarization to higher $G_{\rm c}$. Similarly, in Fig.~\ref{f3_new}(b), increasing RSOC $\lambda_{\rm R}$ lowers the minimum band energy, also increasing $G_{\rm c}$. Circles mark $G_{\rm c}$ for each parameter, defined by $\min[\mathcal{E}_{\vec{k},-}] = \mu$. The actual onset of depolarization, $g_{\rm c}$, occurs slightly after $G_{\rm c}$, since depolarization sets in only when the Fermi surface fully collapses due to EPC-induced energy shifts and band anisotropies. Therefore, $g_{\rm c} \gtrsim G_{\rm c}$.
		
		While the absence of bands or the (an)isotropy of bands near the Fermi energy can be directly inferred from the electronic band structure, additional insights into the effects of the parameters can be obtained by examining the electronic density of states (DOS) in the presence of EPC. In particular, the separation of van Hove singularities along different directions emerges once the spin-resolved band structure becomes anisotropic due to variations in the model parameters. The DOS per unit area is defined as $\mathcal{D(E)} = \frac{1}{(2\pi)^2} \sum_{s=\pm} \int d^2k \, \delta\!\bigl(\mathcal{E} - \mathcal{E}_{{\vec k},s}\bigr)$. In polar coordinates $d^2k = k\,dk\,d\theta$, this becomes
		\begin{align}
			&\mathcal{D(E)} = {} \frac{1}{(2\pi)^2} \sum_{s=\pm} \int_0^{2\pi} d\theta \int_0^\infty k\,dk \delta\!\Bigl(\mathcal{E}  - \tilde{\alpha}_{k,\theta} -s\, \tilde{d}_{k,\theta}\Bigr).
		\end{align}To evaluate the $k$-integral, for each band $s$ we solve the root equation $\mathcal{E} = \tilde{\alpha}_{k,\theta} +s\, \tilde{d}_{k,\theta}$ for positive solutions $k_i$. For a simple isolated root $k_i$ the delta function gives a Jacobian factor, so the contribution of that root is $k_i/(2\pi|\frac{d}{dk}\bigl(\tilde{\alpha}_{k,\theta} +s\, \tilde{d}_{k,\theta}\bigr)\bigl|_{k=k_i}|)$. Therefore, the practical evaluation formula becomes{\small\begin{align}
				\mathcal{D(E)} = \frac{1}{2\pi} \sum_{s=\pm}\sum_{k_i>0}
				\frac{k_i}{\bigg|\tilde{\alpha}'_{k_i}
					+ s\,\frac{\lambda_{\rm R}^2 k_i}{\tilde{d}_{k_i}}
					+ s\,\frac{\bigl(\beta_{k_i} + \tilde g Q_0\bigr)\,\partial_k\beta_k\bigl|_{k=k_i}}{\tilde{d}_{k_i}}\bigg|}\,,
		\end{align}}where $\tilde{\alpha}'_{k_i}=\partial_k\tilde{\alpha}_{k_i}$. Our static-Holstein approach neglects the dynamical electron-phonon self-energy, which in conventional systems with finite $\omega_{\rm p}$ produces kinks in the DOS at $|\mathcal{E}| = \omega_{\rm p}$. In the $\omega_{\rm p} \to 0$ limit, these effects approach the Fermi surface but are absent here, as they stem from the dynamical, not static, self-energy. It should be noted that in the above summation, in order to highlight the anisotropic modifications of the band structure, we treat $k_i$ separately along the $\Gamma \to X$ and $\Gamma \to Y$ paths.\begin{figure}[t]
			\centering
			\includegraphics[width=0.95\linewidth]{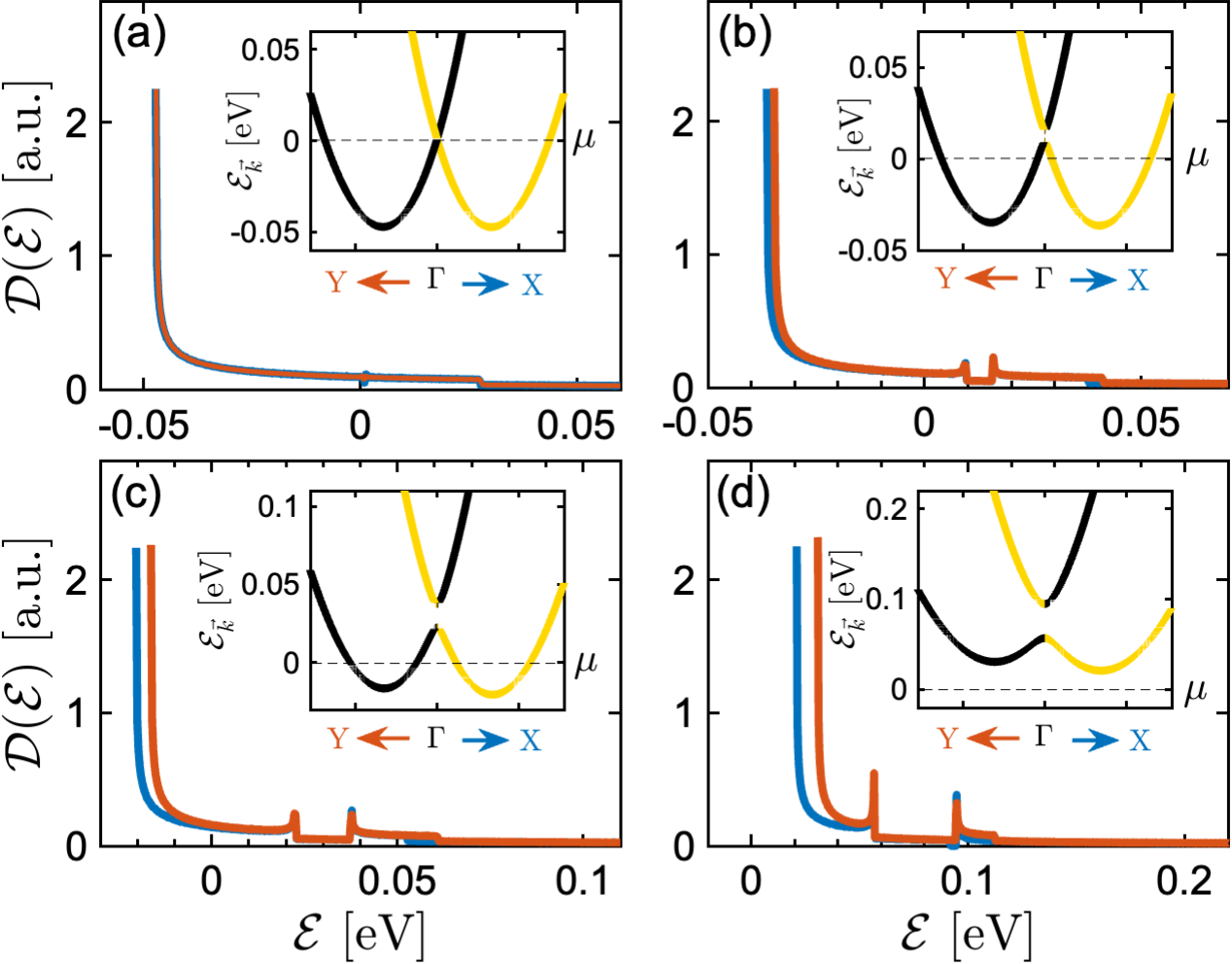}
			\caption{DOS resolved along the 
				$X$ and $Y$ directions for $\beta=0.5$, 
				$\lambda_{\rm R}=0.3$ eV$\cdot$\AA, $\mu=0$, and at different EPCs: 
				(a) $g=0.04$~eV/\AA, (b) $g=0.12$~eV/\AA, 
				(c) $g\approx0.18$~eV/\AA, and (d) $g=0.28$~eV/\AA.  Insets show the corresponding band dispersions along the $\Gamma \to X$ and $\Gamma \to Y$ directions, highlighting the anisotropic 
				reshaping of the bands with increasing $g$. When the bands shift away from the Fermi energy, where the chemical potential is located, the Fermi states are no longer involved due to the absence of available spin states at that energy. As a result, the intraband and interband contributions to the spin Edelstein susceptibility vanish, 
				leading to depolarization. Moreover, the bands become anisotropic with $g$, as reflected in the DOS along different directions.
			}
			\label{f3}
		\end{figure} 
		
		Figure~\ref{f3} shows the DOS $\mathcal{D}(\mathcal{E})$ along the high-symmetry $\Gamma \to X$ and $\Gamma \to Y$ directions for various EPC values $g$. At weak coupling [Fig.~\ref{f3}(a), $g=0.04$ eV/\AA], the band dispersions are nearly symmetric, producing overlapping isotropic DOS peaks corresponding to van Hove singularities from nearly degenerate states. Consequently, the spin Edelstein susceptibilities satisfy $\chi_{\ell j} = -\chi_{j \ell}$, as seen in Fig.~\ref{f2}(a) near $g=0.04$ eV/\AA. Increasing $g$ to $0.12$ eV/\AA\ [Fig.~\ref{f3}(b)] introduces slight anisotropy between $X$ and $Y$, weakly breaking the antisymmetry, $\chi_{\ell j} \neq -\chi_{j \ell}$, consistent with the spin Edelstein responses in Fig.~\ref{f2}(a) near $g=0.12$ eV/\AA.
		
		At $g \approx 0.18$ eV/\AA\, [Fig.~\ref{f3}(c)], the band dispersions become 
		markedly asymmetric associated with more split peaks. The inset shows that the bands 
		become slightly warped: the spin-down dispersion along $X$ shifts downward relative to $Y$. This 
		anisotropy is directly reflected in the DOS, where a double-peak 
		structure begins to form, indicating the splitting of van Hove singularities 
		associated with anisotropic band edges. The threshold EPC, above which permanent depolarization occurs, is $g_{\rm c} \approx 0.25$ eV/\AA. For $g=0.28$ eV/\AA\, [Fig.~\ref{f3}(d)], the DOS peaks are more separated than in the previous case shown in Fig.~\ref{f3}(c). Due to the absence of bands at the Fermi energy~(see the inset showing the spin-resolved band structure in Fig.~\ref{f3}(d)), the intraband and interband contributions to the spin Edelstein susceptibility in Eq.~\eqref{eq_12} are strongly suppressed.
		
		Thereby, the Edelstein depolarization occurs as the phonon-induced energy shift $gQ_0$ elevates the electronic bands above the chemical potential, effectively collapsing the Fermi surface. The detailed dependence of this depolarization on system parameters reflects the intrinsic nonlinearity of the static-Holstein treatment that determines the equilibrium displacement $Q_0$.
	
	In conventional Rashba systems, the Edelstein response is strictly antisymmetric, $\chi_{\ell j} = -\chi_{j\ell}$. In altermagnets, this antisymmetry is generically broken by momentum-dependent spin splitting and its phonon-induced renormalization. The resulting anisotropy can be tuned continuously via the EPC strength, altermagnetic parameter $\beta$, staggered phonon coupling $\tilde g$, Rashba interaction $\lambda_{\rm R}$, and carrier doping, providing a direct route to control direction-dependent spin responses. Importantly, varying these parameters affects the magnitude rather than producing qualitatively different behavior between the $xy$ and $yx$ components. Since depolarization occurs mainly in the dominant transverse susceptibilities, and $\chi_{\ell j} \neq -\chi_{j \ell}$ yet behaves similarly to $-\chi_{j \ell}$, we focus on $-\chi_{xy}$ and analyze its evolution across the parameter space in the presence of EPC.
		
		\subsection{Altermagnetic order effect}
		
		Next, we analyze how altermagnetism affects the persistence of spin Edelstein depolarization, allowing us to contrast the behavior with and without an intrinsic spin-split band structure. Figure~\ref{f4}(a) shows the dependence of the transverse spin Edelstein susceptibility 
		\(-\chi_{xy}/\chi_{0}\) on the EPC strength \(g\) for 
		different values of the staggered lattice potential parameter, or equivalently, the 
		altermagnetic order \(\beta\). For \(\beta = 0\), the system also exhibits a 
		gradual suppression of \(-\chi_{xy}/\chi_{0}\) as \(g\) increases because it is well established that RSOC alone generates a finite spin Edelstein polarization as a direct 
		consequence of inversion-symmetry breaking. Importantly, the depolarization, 
		signaled by the zero of \(-\chi_{xy}/\chi_{0}\), still occurs even in the 
		absence of altermagnetism.\begin{figure}[t]
			\centering
			\includegraphics[width=1\linewidth]{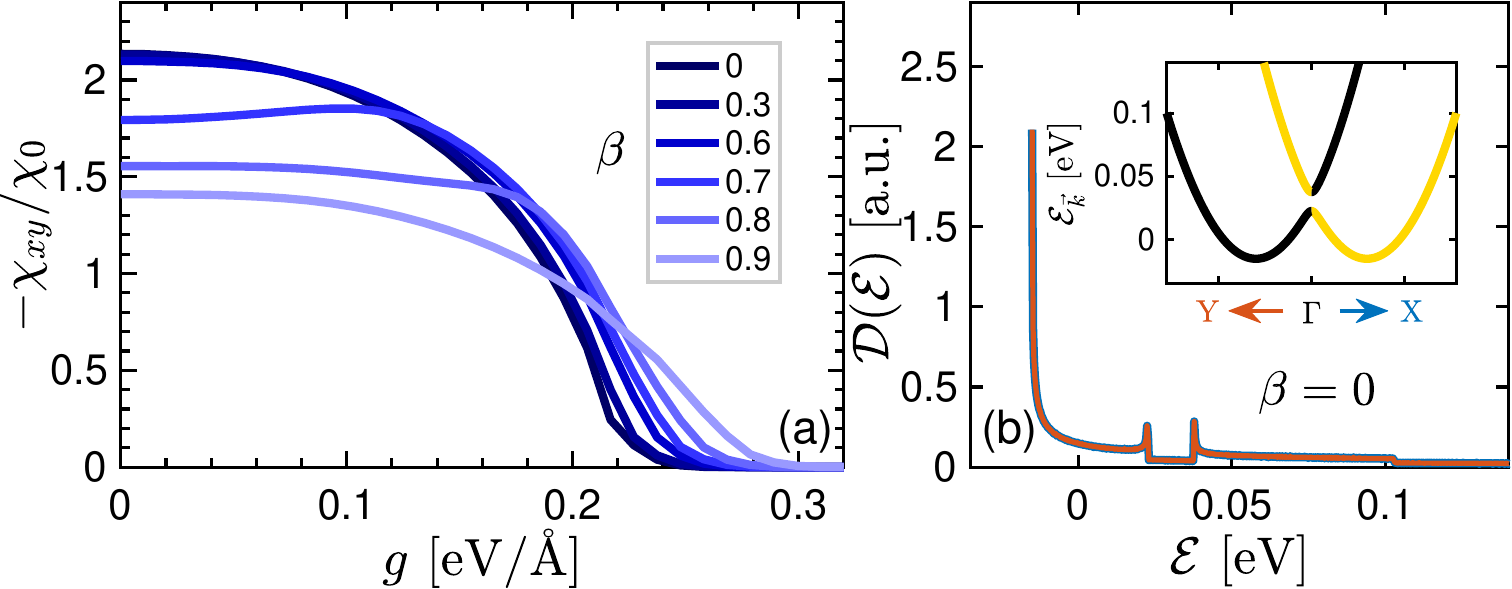}
			\caption{(a) Transverse spin Edelstein susceptibility 
				\(-\chi_{xy}/\chi_{0}\) as a function of EPC
				strength \(g\) for several values of the altermagnetic order \(\beta\). The parameters are set to \(\lambda_{\rm R} = 0.3\) eV$\cdot$\AA\, and \(\mu = 0\). Increasing \(\beta\) slightly shifts the 
				depolarization to larger \(g\), demonstrating strong
				sensitivity of spin Edelstein polarization to lattice asymmetry. 
				(b) Electronic density of states \(\mathcal{D}(\mathcal{E})\) at 
				\(\beta=0\) and $g = 0.2$ eV/\AA, with the inset showing the corresponding band 
				structure \(\mathcal{E}_{\vec{k}}\) along the high-symmetry directions 
				\(Y \leftarrow \Gamma \to X \). Since DOS remains identical along both directions at $\beta = 0$, the 
				resulting finite induced spin (de)polarization is isotropic, which is generally unfavorable for spintronic applications.}
			\label{f4}
		\end{figure}
		
		Although the model inherently predicts an isotropic spin Edelstein polarization at 
		\(\beta = 0\), corresponding to antisymmetric spin susceptibilities, we explicitly 
		validate this behavior by examining the electronic band structure and DOS as in the preceding analyses. Figure~\ref{f4}(b) offers further 
		insight by displaying the electronic DOS for 
		\(\beta = 0\) and $g = 0.2$ eV/\AA, along with the corresponding electronic band structure 
		\(\mathcal{E}_{\vec{k}}\) along the high-symmetry directions 
		\(Y \leftarrow \Gamma \to X\) (inset). 
		
		As illustrated, the DOS along both the 
		\(X\) and \(Y\) directions exhibits identical features at \(\beta = 0\), confirming 
		that in the absence of altermagnetic order, the induced spin Edelstein (de)polarization 
		remains highly isotropic. By contrast, the introduction of a finite altermagnetic order parameter \(\beta\),
		breaks this isotropy by inducing anisotropic spin-splitting in the band structure, 
		thereby rendering the spin Edelstein response directionally dependent. This 
		anisotropy is a crucial attribute for spintronic applications, where control over 
		the spin polarization along specific crystallographic directions is often desired.
		
		As \(\beta\) increases, the induced spin polarization decreases and the depolarization point shifts slightly toward larger threshold EPC strengths \(g_{\rm c}\), indicating that the altermagnetic strength delays the onset of depolarization. This was already confirmed by Fig.~\ref{f3_new}(a).
		Physically, this can be attributed to 
		the enhanced band structure induced by the combined effects of EPC and altermagnetic 
		order, which strongly suppresses the intraband and interband contributions to the 
		susceptibility.
		
		\subsection{Phonon-altermagnetic order coupling effect}
		As the next investigation, we examine how EPC to the altermagnetic order~(\redd{due to $C_4 \mathcal{T}$ symmetry breaking}), $\tilde{g}/g$, affects the persistence of spin Edelstein depolarization. It is observed, in Fig.~\ref{f5}(a), that for $\tilde{g}/g < 1$, the rate of polarization suppression depends only weakly on $\tilde{g}/g$. Larger values of the ratio (but still below 1) lead to a slower increase in $-\chi_{xy}/\chi_{0}$ and slightly shift the zero to larger $g$, implying 
		that a stronger coupling to the staggered potential slightly enhances the 
		susceptibility renormalization. 
		
		At $\tilde{g}/g = 1$, there is a strong shift of the bands, such that a much larger EPC is required to suppress the spin Edelstein polarization, as the intraband and interband transitions still contribute significantly to the susceptibility. This, in turn, means that stronger $\tilde{g}/g$ stabilizes the transverse polarization 
		more efficiently.\begin{figure}[t]
			\centering
			\includegraphics[width=1\linewidth]{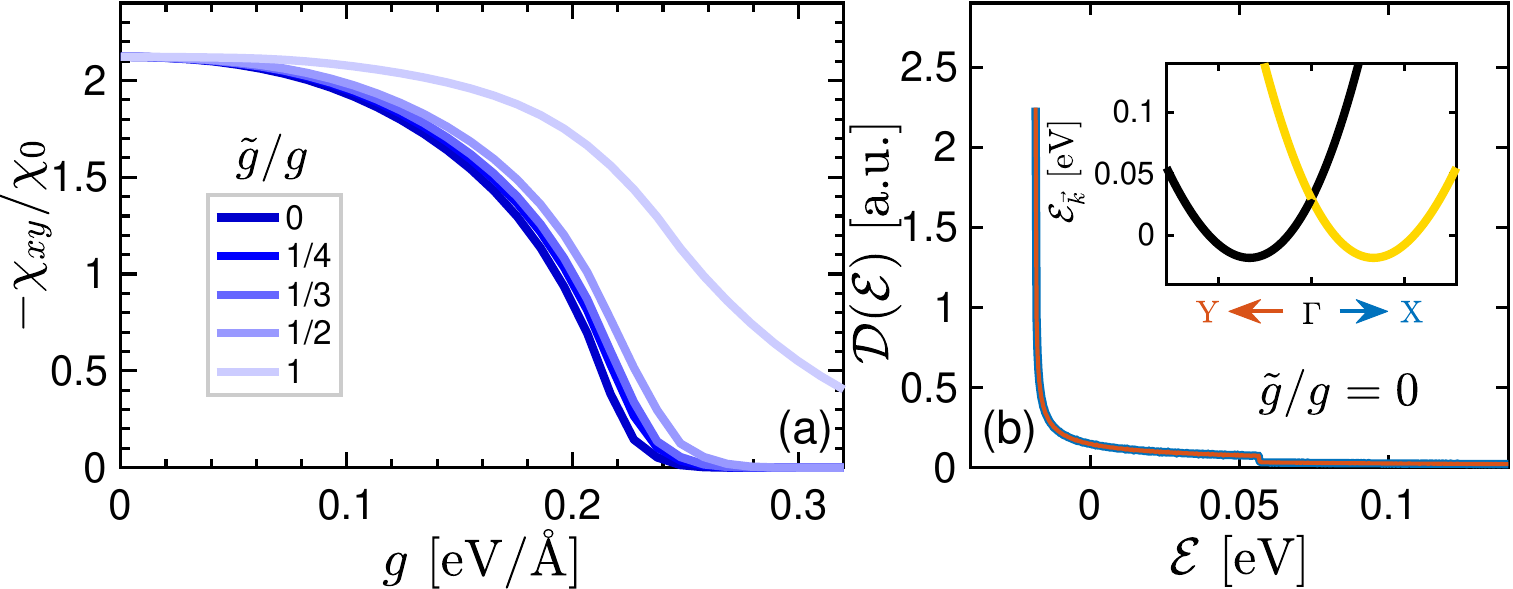}
			\caption{(a) Transverse susceptibility $-\chi_{xy}/\chi_{0}$ as a 
				function of the EPC strength $g$ for different phonon coupling to  the 
				staggered lattice potential~(\redd{$C_4 \mathcal{T}$ symmetry breaking}), $\tilde{g}/g$. The parameters are fixed at 
				$\beta=0.5$, $\lambda_{\rm R}=0.3$ eV $\cdot$\AA, and $\mu=0$. 
				The curves show that increasing $\tilde{g}/g$ slightly shifts the zero 
				of $-\chi_{xy}$ to larger $g$, leading to a slower decay 
				of the transverse response. (b) Electronic DOS at \(\tilde{g}/g = 0\) and $g = 0.2$ eV/\AA, with the inset showing the corresponding band structure \(\mathcal{E}_{\vec{k}}\) near the $\Gamma$ point. As previously noted, the parameters $\tilde{g}$ and $\beta$ are interdependent, with the activation or deactivation of one directly affecting the presence of the other. Thus, at \(\tilde{g}/g = 0\), the DOS is also identical along both directions, resulting in an isotropic induced spin (de)polarization.
			}
			\label{f5}
		\end{figure}
		
		We further highlight that in the absence of explicit EPC to the staggered potential, depolarization still occurs. However, in Fig.~\ref{f5}(b) at $g = 0.2$ eV/\AA, our DOS analysis and band structure (inset of Fig.~\ref{f5}(b)) reveal that the induced polarization and depolarization remain isotropic at $\tilde{g}/g = 0$, since the DOS is identical along the $X$ and $Y$ directions of the Brillouin zone. The parameters $\tilde{g}$ and $\beta$ are then intrinsically linked, such that changes in one necessarily influence the other, reflecting their mutual dependence within the system. Similar to the case of $\beta = 0$, such isotropy is undesirable for spintronics, where directional control of spin responses is essential.
		
		\subsection{Rashba spin-orbit coupling effect}
		We now investigate how electrostatic gating influences the spin (de)polarization in the presence of EPC. As explained earlier, applying a gate induces RSOC. Figure~\ref{f6}(a) shows the spin Edelstein susceptibility 
		\(-\chi_{xy}/\chi_{0}\) as a function of the EPC strength 
		\(g\) for several values of the RSOC strength \(\lambda_{\rm R}\). 
		The calculations are performed for fixed \(\beta = 0.5\) and chemical potential \(\mu = 0\).  
		Several important features emerge from this figure. First, at $\lambda_{\rm R}=0$, spin polarization vanishes since inversion symmetry remains unbroken.
		Second, as the EPC strength increases, a gradual suppression of 
		\(-\chi_{xy}/\chi_{0}\) is observed for $\lambda_{\rm R} < 0.4$ eV$\cdot$\AA, consistent with the general response 
		trends reported in earlier analyses. Decreasing $\lambda_{\rm R}$ shift the depolarization to occur 
		at smaller threshold EPC strengths. For example, decreasing 
		\(\lambda_{\rm R}\) from \(0.3\) eV$\cdot$\AA\, to 
		\(0.1\) eV$\cdot$\AA\, reduces the corresponding \(g_{\rm c}\) from $0.25$ eV/\AA\, to 0.13 eV/\AA, indicating that stronger spin-orbit coupling facilitates the onset of 
		spin Edelstein depolarization at stronger EPC.\begin{figure}[t]
			\centering
			\includegraphics[width=1\linewidth]{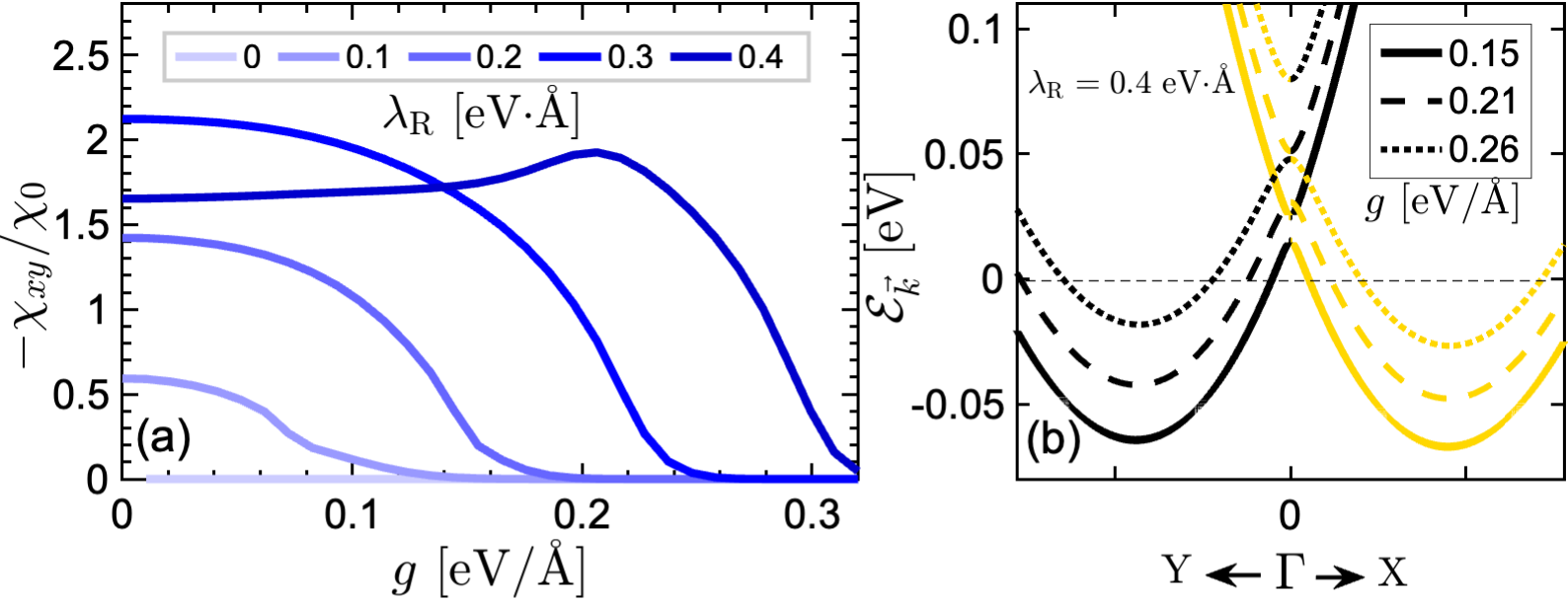}
			\caption{(a) Transverse spin Edelstein susceptibility 
				\(-\chi_{xy}/\chi_{0}\) as a function of the EPC strength \(g\) for different values of the Rashba coupling 
				\(\lambda_{\rm R}\), with parameters 
				\(\beta = 0.5\), and chemical potential 
				\(\mu = 0\). Increasing $\lambda_{\rm R}$ delays the onset of the depolarization. This illustrates the competing role of gating-induced RSOC, which, while useful for tuning spin textures, simultaneously drives the system toward a slower loss of polarization. (b) Spin-resolved band structure near \(g \approx 0.21\) eV/\AA, where \(-\chi_{xy}/\chi_{0}\) exhibits nonmonotonic behavior at \(\lambda_{\rm R} = 0.4\) eV$\cdot$\AA. The peak in \(-\chi_{xy}/\chi_{0}\) arises from the increased number of Fermi states participating in intraband and interband transitions at this EPC.
			}
			\label{f6}
		\end{figure}   
		
		Physically, an increase in the RSOC enhances both the spin-momentum locking and the effective spin splitting of the 
		bands, which in turn tends to preserve finite spin states at the Fermi surface. This makes the current-induced spin polarization more 
		robust against EPC-induced band renormalization, thereby delaying the suppression of the current-induced spin 
		polarization.
		
		Notably, for larger values of the Rashba coupling, \(\lambda_{\rm R} \geq 0.4\) eV$\cdot$\AA, the induced spin polarization exhibits a non-monotonic behavior: it is slightly enhanced for EPC strengths up to \(g \approx 0.21\) eV/\AA, followed by a gradual decrease for \(g > 0.21\) eV/\AA. To elucidate this behavior, we analyze the spin-resolved band structure near this threshold EPC strength in Fig.~\ref{f6}(b). As soon as the EPC becomes sufficiently strong to shift the spin-band edge to the Fermi level, the spin susceptibility reaches its maximum. This peak arises because the involved spin states are largest at this point, leading to the strongest contributions from both intraband and interband transitions. Beyond this regime, as the number of available states participating in these transitions decreases with increasing \(g\), the susceptibility correspondingly decreases.
		
		\subsection{Electron and hole doping effects}
		Finally, we examine the case of the doping effect, in which the chemical potential is shifted away from the zero Fermi energy. Figure~\ref{f7} demonstrates the influence of the chemical 
		potential $\mu$ on the transverse spin Edelstein susceptibility 
		$-\chi_{xy}/\chi_{0}$ as the EPC strength $g$ is varied. \begin{figure}[t]
			\centering
			\includegraphics[width=1\linewidth]{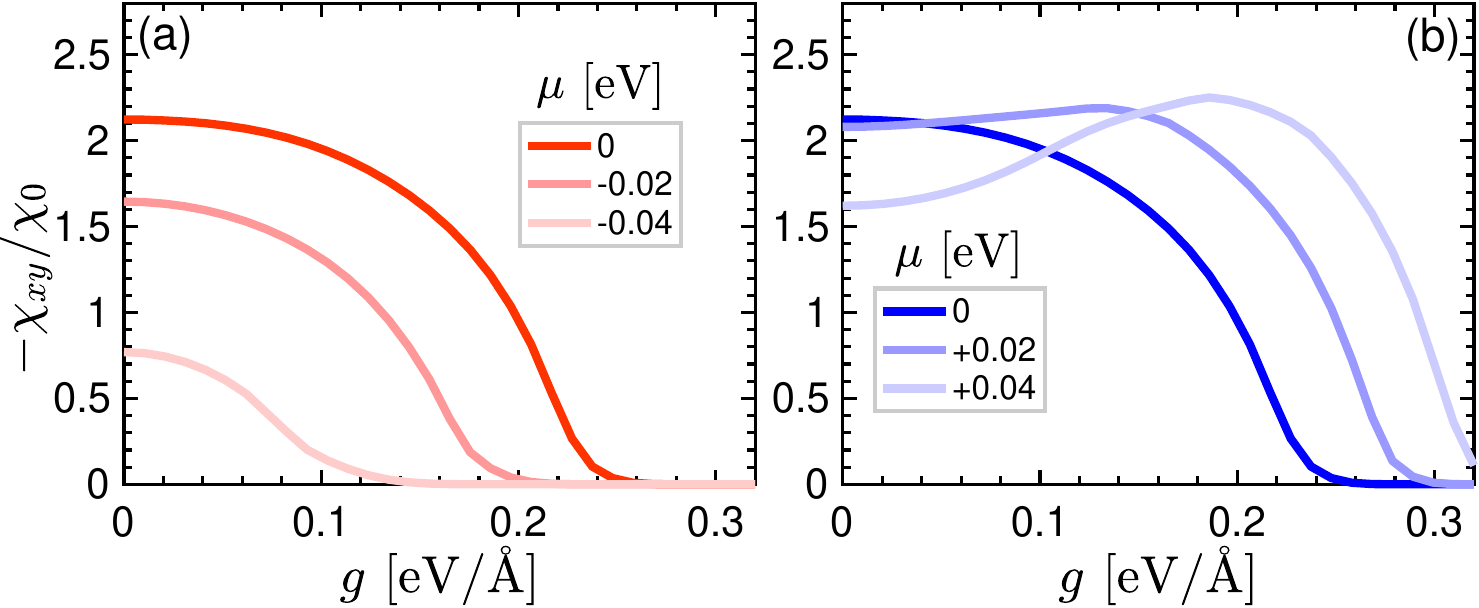}
			\caption{Transverse spin Edelstein susceptibility 
				$-\chi_{xy}/\chi_{0}$ as a function of the EPC
				strength $g$, for different values of the chemical potential $\mu$ corresponding to (a) hole and (b) electron doping, respectively. 
				Parameters are fixed to $\beta=0.5$, and
				$\lambda_{\rm R}=0.3$ eV$\cdot$\AA. Spin Edelstein depolarization occurs at progressively smaller or larger EPC strengths as the doping level shifts downward or upward from the Fermi energy toward the spin-split bands, respectively.
			}
			\label{f7}
		\end{figure}
		
		When the chemical potential is shifted below the band edge 
		($\mu=-0.02$ and $-0.04~\mathrm{eV}$) due to hole doping, Fig.~\ref{f7}(a), the magnitude of $-\chi_{xy}$ is reduced 
		compared to the $\mu=0$ case, indicating that fewer carriers at the Fermi level 
		participate in the Edelstein response. Interestingly, as $\mu$ decreases, depolarization occurs at smaller EPC strengths. This again originates from the reshaping of the Fermi 
		surface in the presence of both Rashba coupling and electron-phonon 
		renormalization.
		
		For positive chemical potentials (\(\mu = +0.02\) and \(+0.04~\mathrm{eV}\)), corresponding to electron doping, the behavior shown in Fig.~\ref{f7}(b) differs significantly. The susceptibility curve exhibits a nonmonotonic dependence on \(g\), characterized by an intermediate enhancement before eventually decaying. The peak in the susceptibility can be understood in the same way as in Fig.~\ref{f6}(b): for \(\mu > 0\) and \(g > 0.15\) eV/\AA, an increasing number of states participate in intraband and interband transitions, leading to a temporary increase in susceptibility. Beyond this point, as the number of available states diminishes, the susceptibility begins to decrease. Thus, a strong particle-hole asymmetry appears in the Edelstein response.
		
		\section{Experimental perspective}\label{s5}
		Before concluding, we comment on the experimental feasibility of our predictions. A promising route to experimentally probe the proposed effects is to interface a 2D \(d\)-wave altermagnet---such as thin films of KV\(_2\)Se\(_2\)O~\cite{Jiang2025}, RbV\(_2\)Te\(_2\)O~\cite{Zhang2025}, RuO\(_2\) (noting that the altermagnetic character of RuO\(_2\) remains under active investigation)~\cite{Feng2022,doi:10.1126/sciadv.aaz8809,PhysRevLett.128.197202,weber2024opticalexcitationspinpolarization}, and \(\kappa\)-Cl~\cite{Naka2019}---with an appropriate substrate at low temperatures to induce lattice vibrations in the altermagnet layer. Crucially, since the underlying phonon-mediated mechanism requires only a linear coupling between lattice displacements and spin states, the proposed effect should be realizable across a broad class of altermagnetic thin films. Electrostatic control over the RSOC can be achieved using top and bottom gate electrodes 
		separated by high-quality dielectric layers. Complementary to phonon generation and gating, angle-resolved photoemission spectroscopy enables 
		direct visualization of spin textures, band splittings, and symmetry-breaking phenomena~\cite{10.1063/5.0151859}. 
		Additionally, the Edelstein effect can be probed in Hall-bar geometries through 
		electrical transport measurements. 
		
		We note, however, that our analysis relies on a low-energy static-Holstein model with a simplified, non-dispersive phonon mode. While it captures the essential physics 
		of EPC-driven spin (de)polarization, effects arising from full-band structures, dispersive phonons, 
		or beyond static-Holstein interactions may quantitatively or qualitatively modify the results. 
		Accordingly, the present findings should be regarded as minimal-model predictions.
		
		\section{Summary and outlook}\label{s6}
		In this work, we have examined how slow lattice vibrations shape spin responses in $d$-wave altermagnets, systems that combine vanishing net magnetization with spin-split electronic bands. By formulating the problem within a Rashba continuum model coupled to slow phonons  through a Holstein-type interaction, and treating the EPC at the static-Holstein level, we have tracked the evolution of the Edelstein effect over a broad range of parameters. The Kubo formalism provided access to distinguish the specific contributions of intraband and interband processes. \redd{In our framework, EPC plays a dual role. It renormalizes the electronic dispersion through an effective shift of the chemical potential and, when tied to a static lattice distortion—such as a piezomagnetically active strain—acts as a symmetry-breaking field that lifts the $C_4\mathcal{T}$ symmetry. Thus, EPC is not merely a passive renormalization effect but an active mechanism that reshapes the electronic structure and enables the symmetry-driven phenomena discussed here.}
		
		A central result of our study is that increasing EPC gradually suppresses the Edelstein spin polarization, leading to a threshold beyond which the spin polarization collapses to zero, signaling a qualitative restructuring of the spin channels. In this regime, both intraband and interband transitions at the Fermi energy vanish, and the effective Fermi surface, required for the Edelstein response, disappears. The (an)isotropic character of the depolarization strongly depends on altermagnetism: without it, the effect is isotropic, while its interplay with altermagnetism induces pronounced anisotropy and breaks the usual antisymmetry of the spin susceptibilities, highlighting altermagnetism as key for directionally selective spin responses. Our parameter-space analysis shows that the onset of depolarization is tunable: the threshold EPC shifts with staggered lattice coupling, Rashba interaction, and carrier doping. Moving the chemical potential away from zero modifies the threshold, with hole doping accelerating depolarization and electron doping delaying it by enhancing the susceptibility.

		As future directions, first, exploring the interplay of phonon-induced depolarization with other spin-orbit interactions, such as Dresselhaus or proximity-induced effects, could reveal richer anisotropic spin responses. Second, extending these studies to multilayer altermagnets or heterostructures may enable enhanced tunability via interlayer coupling and interface engineering. Third, time-resolved experiments, including ultrafast optical or terahertz pump-probe techniques, could probe the dynamical evolution of the Edelstein effect under controlled phonon excitations. Fourth, combining phonon engineering with strain, gating, or substrate manipulation could allow precise control over spin polarization for optimized spintronic devices. Finally, applying these concepts to other materials, such as topological insulators, 2D magnets, or Rashba systems with strong correlations, may reveal universal features of phonon-controlled spin responses. We leave these topics for future work.
		
		\section*{Acknowledgments}
		M. Y. gratefully acknowledges useful discussions with Ulrich Z\"{u}licke. M. Y. and J. K. F. were supported by the Department of Energy, Office of Basic Energy Sciences, Division of Materials Sciences and Engineering under Contract No. DE-FG02-08ER46542 for the formal developments, the numerical work, and the writing of the manuscript. J. K. F. was also supported by the McDevitt bequest at Georgetown University. J. L. was supported by the Research Council of Norway through its Centres of Excellence funding scheme Grant No. 262633 and Grant No. 353894 for writing of the manuscript.
		
		\section*{Data availability}
		The data supporting the findings of this article are openly available in the Zenodo database~\cite{Zenodo}.
	}
		\appendix
		{\allowdisplaybreaks
			\section{Derivation of the phonon-dressed low-energy Hamiltonian}\label{ap1}
				In this appendix, we explicitly perform the projection of the lattice electron-phonon Hamiltonian $H_{\mathrm{e\text{-}p}} = \sum_{i,\nu} Q^\nu_i\, c^\dagger_i \hat{O}_\nu c_i$, given in Sec.~\ref{s2b}, onto our low-energy Rashba-altermagnetic subspace in the adiabatic phonon regime, where we restrict to the lowest phonon mode $\nu=0$ at
		$\vec q=\vec 0$, so that the phonon coordinate is uniform in space, $Q_0(\vec R_i) \equiv Q_0$. The lattice electron-phonon Hamiltonian then reduces to
		\begin{align}\label{eq:Hepphi0}
			H_{\mathrm{e\text{-}p}}^{(0)}
			=
			Q_0
			\sum_i
			\psi_i^\dagger
			\left(
			g_A^x e_A^{(0)x}
			+
			g_B^y e_B^{(0)y}
			\right)
			\psi_i .
		\end{align}
		For the correlated displacement pattern $e_A^{(0)x}=1$ and $e_B^{(0)y}=1$, the electronic operator associated with the frozen phonon mode is $\hat O_0
			=
			g_A^x
			+
			g_B^y$.
		Up to this point, no approximation has been made. Near the $\Gamma$ point, the low-energy electronic Hilbert space is spanned
		by two bands $s=\pm$ characterized by spin-sublattice (two sublattices $A$ and $B$) locking. Explicitly,
		the Bloch spinors may be written as $|u_+\rangle
		= (|A,\uparrow\rangle,0)^T$ and $|u_-\rangle
		= (0,|B,\downarrow\rangle)^T$.	These states are orthonormal and satisfy $\sigma_z |u_\pm\rangle = \pm |u_\pm\rangle$.
		
		The projection of $\hat O_0$ onto the low-energy subspace is
		\begin{align}\label{eq:O0proj}
			\hat O_0
			=
			\sum_{s,s'=\pm}
			|u_{s'}\rangle
			\langle u_{s'}|
			\left(
			g_A^x+g_B^y
			\right)
			|u_s\rangle
			\langle u_s| .
		\end{align}Since $g_A^x$ acts only on sublattice $A$ with spin-up and $g_B^y$ only on
		sublattice $B$ with spin-down, the matrix elements are diagonal, $\langle u_{s'}|\hat O_0|u_s\rangle
			=
			\delta_{ss'} g_\uparrow^x$ for $s=+$ and $\langle u_{s'}|\hat O_0|u_s\rangle
			=
			\delta_{ss'} g_\downarrow^x$ for $s=-$, where
		\begin{align}
			g_\uparrow^x \equiv \langle A,\uparrow|g_A^x|A,\uparrow\rangle,
			\qquad
			g_\downarrow^y \equiv \langle B,\downarrow|g_B^y|B,\downarrow\rangle .
		\end{align}
		Thus, in the $\{|u_+\rangle,|u_-\rangle\}$ basis, we obtain
		\begin{align}\label{eq:O0matrix}
			\hat O_0
			=
			\begin{pmatrix}
				g_\uparrow^x & 0 \\
				0 & g_\downarrow^y
			\end{pmatrix} = \frac{g_\uparrow^x}{2} \left(\sigma_0 + \sigma_z\right) + \frac{g_\downarrow^y}{2} \left(\sigma_0 - \sigma_z\right),
		\end{align}leading to $\hat O_0
			=
			\frac{g_\uparrow^x+g_\downarrow^y}{2}\,\sigma_0
			+
			\frac{g_\uparrow^x-g_\downarrow^y}{2}\,\sigma_z$. Thus, the resulting low-energy electron-phonon Hamiltonian reads
		\begin{align}\label{eq:Heplowfinal}
			H_{\mathrm{e\text{-}p}}
			=
			Q_0
			\sum_i
			c_i^\dagger
			\left(			\frac{g_\uparrow^x+g_\downarrow^y}{2}\,\sigma_0
			+
			\frac{g_\uparrow^x-g_\downarrow^y}{2}\,\sigma_z
			\right)	c_i .
		\end{align}\redd{In the pristine lattice, the two sublattices $A$ and $B$ are related by a lattice symmetry operation (rotation or mirror, depending on lattice type) followed by time reversal. The degeneracy at $\Gamma$ is therefore protected by the $C_4 \mathcal{T}$ symmetry. A finite staggered coupling $g_\uparrow^x \neq g_\downarrow^y$ requires a lattice distortion that breaks $C_4 \mathcal{T}$ symmetry. Appropriate symmetry-lowering piezomagnetism strain patterns generate $g_\uparrow^x \neq g_\downarrow^y$~\cite{PhysRevMaterials.8.L041402,doi:10.7566/JPSJ.94.083702,PhysRevB.110.144421,10.1063/5.0277631,bell2026orbitalpiezomagneticpolarizabilitypure,10.1063/5.0242426,doi:10.7566/JPSJ.94.063704,doi:10.7566/JPSJ.94.083702,khodas2025tuningaltermagnetismstrain}.} 
	}
	\bibliography{bib.bib}
\end{document}